\documentclass[a4paper,aps,showpacs]{revtex4}
\usepackage[reqno]{amsmath}
\usepackage{epsf,amsfonts,amssymb,dcolumn}
\newcommand{\ebox}[2]{\epsfxsize=#1 \epsfbox[10 30 560 590]{#2}}

\newcommand{\Dslash}{/\hspace{-1.5ex}D}
\newcommand{\Dslashbak}{\overleftarrow{\Dslash}}
\newcommand{\dslash}{\not\!\partial}
\newcommand{\pslash}{\not\!p}

\newcommand{\kslash}{\not\!k}
\newcommand{\half}{\frac{1}{2}}
\newcommand{\bra}{\langle}
\newcommand{\ket}{\rangle}
\newcommand{\Tr}{\operatorname{tr}}

\newcommand{\order}{\mathcal{O}}

\newcommand{\khat}{\hat{k}}
\newcommand{\psibar}{\overline{\psi}}
\newcommand{\Bi}{B'_I}
\newcommand{\zz}{Z^{(0)}}
\newcommand{\dmz}{\Delta M^{(0)}}
\newcommand{\sss}{\scriptscriptstyle}
\newcommand{\muv}{m_{\sss\rm UV}}
\newcommand{\MIR}{M_{\sss\rm IR}}

\newcommand{\err}[2]{\raisebox{-0.4ex}
{$\stackrel{\scriptstyle +#1}{\scriptstyle -#2}$}}

\begin{document}

\preprint{ADP-00-40/T423 \hskip0.5cm
          DESY 00-099}

\title{Quark Propagator in Landau Gauge}

\author{
Jon Ivar Skullerud}
\email{jonivar@mail.desy.de}
\homepage{http://www.bigfoot.com/~jonivar/}
\affiliation{CSSM and Department of Physics and Mathematical Physics, \\
Adelaide University, Australia 5005\\
and \\
DESY Theory Group, Notkestra{\ss}e 85, D--22603 Hamburg, Germany}
\author{Anthony G.\ Williams}
\email{awilliam@physics.adelaide.edu.au}
\homepage{http://www.physics.adelaide.edu.au/cssm/}
\affiliation{CSSM and Department of Physics and Mathematical Physics, \\
Adelaide University, Australia 5005}

\begin{abstract}
The Landau gauge quark propagator in momentum space is investigated
using the $\order(a)$-improved Sheikholeslami--Wohlert (SW) quark
action with a tree-level mean-field improved coefficient $c_{sw}$.  We
have studied the unimproved definition of the quark propagator, as
well as two different tree-level
$\order(a)$-improved propagators.  The ultraviolet behavior of the free
lattice propagator is studied for each of these in order to establish
which of them provides the most reliable description of the quark
propagator up to the medium momentum regime.  A general method of
tree-level correction is introduced.  This exploits asymptotic
freedom and removes much of the trivial lattice artifacts at medium
to high momenta.  We obtain results for
the quark propagator which are qualitatively similar to those
typically used in quark models.  A simple extrapolation of the infrared
quark mass $M(p^2=0)$ to the chiral limit gives $298\pm8\pm30$~MeV,
which is consistent with phenomenological expectations.
\end{abstract}


\maketitle

\section{Introduction}
\label{Sec:intro}

The quark propagator is one of the fundamental quantities in QCD.  By
studying the momentum-dependent quark mass, obtained from the scalar part
of the inverse quark propagator, we can gain valuable insight into the
mechanism of chiral symmetry breaking and its momentum dependence.
The quark propagator is also used extensively as an input in
Dyson--Schwinger \cite{Roberts:1994dr} based model calculations of
hadronic matrix elements \cite{Maris:2000sk,Maris:2000bh}.  Hence a
lattice calculation of the quark propagator would enable us to check the
validity of the models used in these calulations.
There have been several recent studies of the quark propagator on
a finer lattice with the
aim of obtaining the light quark masses and renormalization constants
\cite{Becirevic:2000kb}.  Here we will focus more on the infrared and
medium-momentum regime and extend some earlier preliminary
work~\cite{Skullerud:2000gv}.
For comparison with the present studies, some results for the
quark mass function using
Kogut-Susskind fermions have recently been reported~\cite{Aoki:99}.  

The study of the quark propagator on the lattice is complicated by the
explicit chiral symmetry breaking in the Wilson fermion action, and
also by finite lattice spacing effects, which are large compared to
those in the pure gauge sector \cite{Leinweber:1999uu}.  For the gluon
sector, on the contrary, one can achieve reliable results even with
very coarse lattices using $\order(a^2)$-improved actions together
with mean-field improvement.  These studies have
shown, for example, that in Landau gauge the gluon propagator is
enhanced at intermediate momenta and suppressed in the infrared to the
point where it is almost certainly infrared finite~\cite{Bonnet:2000kw}.

Perturbation theory in a covariant gauge has a gauge-fixing
parameter $\xi$, which corresponds to the width of the Gaussian
average over the auxiliary field $c(x)$ in the gauge fixing condition
$\partial_\mu A_\mu(x)=c(x)$.  
The choice of Landau gauge, i.e., $\xi=0$, corresponds to the zero width
case, i.e., $\partial_\mu A_\mu(x)=0$, which is the Lorentz gauge-fixing
condition.  Hence ``covariant gauges'' are actually Gaussian weighted
averages over
generalizations of the Lorentz gauge fixing condition.  On the lattice,
Landau gauge means that we have imposed the Lorentz gauge condition
by finding a local minimum of the appropriate gauge-fixing functional
\cite{Leinweber:1999uu}.
As for the previously cited studies of the gluon propagator, the
work reported here is done in the quenched approximation and without
attempting to avoid Gribov copies, e.g., without attempting to project
onto the fundamental modular region.  Of course, in our finite ensemble
of gauge field configurations no two Landau-gauge configurations
will ever be Gribov copies of each other.  However, the Landau gauge
configurations will not be samples from a single connected manifold
such as the fundamental modular region.  This is an interesting
area for future study.

In a covariant gauge in the continuum the renormalized
Euclidean space quark propagator must have the form
\begin{equation}
S(\mu;p) = \frac{Z(\mu;p^2)}{i\pslash + M(p^2)}\equiv 
     \frac{1}{i\pslash A(\mu;p^2) + B(\mu;p^2)} \;,
\label{Eq:generic-prop}
\end{equation}
where we see that $Z(\mu;p^2)\equiv 1/A(\mu;p^2)$ and
$M(p^2)\equiv B(\mu;p^2)/A(\mu;p^2)$.  The renormalization point is
denoted by $\mu$ and since we are interested in defining
{\em nonperturbative} renormalization we use the standard momentum subtraction
scheme (MOM), which has the renormalization point boundary
conditions
\begin{equation}
Z(\mu,\mu^2)=1 \hskip1cm {\rm and} \hskip1cm M(\mu^2) = m(\mu)\; .
\label{Eq:ren_pt}
\end{equation}
At sufficiently large $\mu$ in an asymptotically free theory like QCD
the effects of dynamical chiral symetry breaking become small
and $m(\mu)$ becomes the usual explicit chiral symmetry breaking
running quark mass.  At large Euclidean momentum scales (i.e., large $\mu$)
the procedure for relating the parameters of the MOM scheme to the
popular {\em perturbative} renormalization schemes 
(i.e., MS or $\overline{\rm MS}$) is well known.

The renormalizability of QCD implies that the bare propagator is related
to the renormalized one through the quark wavefunction renormalization
constant $Z_2$:
\begin{equation}
S^{\rm bare}(a;p) = Z_2(\mu;a) S(\mu;p) \; ,
\label{Eq:bare_to_ren}
\end{equation}
where $a$ denotes the regularization parameter in some regularization
scheme (such as the lattice or dimensional regularization).
In a renormalizable theory the renormalized
quantities become independent of the regularization parameter in the limit
that it is removed (i.e., $a\to 0$ on the lattice or $\epsilon\to 0$ in a
dimensional regularization scheme),
while holding the renomalization point boundary conditions fixed
in Eq.~(\ref{Eq:ren_pt}).  It immediately follows from the renormalization
point independence of the LHS of Eq.~(\ref{Eq:bare_to_ren}) that
for sufficently small $a$ (i.e., in the {\em scaling region}) we have
\begin{equation}
\frac{Z_2(\mu;a)}{Z_2(\mu';a)}=\frac{Z(\mu',p^2)}{Z(\mu,p^2)}  \hskip1cm
  {\rm and} \hskip1cm M(p^2)\equiv M(\mu;p^2)=M(\mu';p^2)
\label{Eq:mu_dependence}
\end{equation}
for all $p^2$.  Hence the mass function must be renormalization
point independent and a change of renormalization point is just an
overall rescaling of $Z(\mu;p^2)$ by a momentum-independent constant, i.e.,
the LHS of the first equality in Eq.~(\ref{Eq:mu_dependence}).
Hence, once the momentum-dependent renormalized propagator is known at
one $\mu$ for all $p$, then it is immediately known for all $\mu$. 
We can evaluate the constant needed to rescale $Z(\mu;p^2)$ to
$Z(\mu';p^2)$ by evaluating
Eq.~(\ref{Eq:mu_dependence}) at $p^2={\mu'}^2$ and using
$Z(\mu';{\mu'}^2)=1$   [i.e., Eq.~(\ref{Eq:ren_pt})] to give
\begin{equation}
\frac{Z_2(\mu';a)}{Z_2(\mu;a)}=Z(\mu,{\mu'}^2) \; .
\end{equation}

Perturbative QCD chooses a renormalization scale $\mu$ close to the
momentum scale characterizing the particular process of interest, e.g.,
$\mu^2\sim Q^2$ in deep inelastic scattering.  This choice is made to
ensure that perturbation theory will converge as rapidly as possible
for the process of interest.
The quark propagator used in such calculations is then $S(\mu;p)$
for $p^2$ near $\mu^2$, i.e., 
$S^{\rm perturb}(\mu;p)\equiv 1/[i\pslash+m(\mu)]$.

The tree-level quark propagator is the bare (i.e., regularized)
quark propagator {\em in the absence of interactions}, i.e.,
\begin{equation}
S^{(0)}(a;p) = \frac{1}{i\pslash + m^0(a)} \;,
\end{equation}
where $m^0(a)$ is the bare quark mass.  When the interactions with the gluon
field are turned on then
\begin{equation}
S^{(0)}(p) \to S^{\rm bare}(a;p) = Z_2(\mu;a)S(\mu;p)  \;.
\label{Eq:S_relns}
\end{equation}
So we see that in the {\em scaling region}
(i.e., for sufficiently small $a$) the
measure of nonperturbative physics is the deviation of
$Z(\mu;p^2)$ from 1 and the difference of $M(p^2)$ from the renormalized
quark mass $m(\mu)$.  As already mentioned, for sufficiently large
$\mu$, $m(\mu)$ becomes the running mass at the renormalization
point $\mu$, which is the basis of the studies performed in
Ref.~\cite{Becirevic:2000kb}.

The purpose of the work reported here is to extract the full
$Z(\mu;p^2)$ and $M(p^2)$ directly from a lattice calculation
of the bare quark propagator $S^{\rm bare}(a;p)$.  It is of course
$S^{\rm bare}(a;p)$ that is calculated on the lattice. In reality
we do not have the convenience of
having an arbitrarily small $a$, rather we are faced with a lattice
spacing which introduces lattice artifacts at medium to high momenta.
In order to simplify the presentation of the data we will not explicitly
introduce a renormalization point.  Rather we will introduce for
convenience the renormalization-point-independent combination
\begin{equation}
Z(p^2)\equiv Z_2(\mu;a) Z(\mu;p^2) \; .
\end{equation}
The regularization parameter dependence (i.e., the $a$-dependence)
of $Z(p^2)$ is not indicated for brevity, but is to be understood.
We will develop a procedure for tree-level correction of the
lattice artifacts in order to minimize their effect.

The structure of the remainder of this paper is as follows: In
Sec.~\ref{Sec:improved_quarks} we describe the various $\order(a)$
improved quark actions and propagators that we will study.  In
Sec.~\ref{Sec:mom_space_quarks} we introduce our notation for the
propagator in momentum space and derive the tree-level expressions
appropriate for the actions that we consider.
In Sec.~\ref{Sec:analysis} we describe our methods to minimize lattice
artifacts, including our tree-level correction scheme.
Section~\ref{Sec:results} contains our numerical results:  In
Sec.~\ref{Sec:unsubt_data} we present the data for the propagator
without the tree-level correction;  the effect of the tree-level
correction is shown in Sec.~\ref{Sec:tree_level_subtr_data};  in
Secs.~\ref{Sec:model_fits} and \ref{Sec:IR_mass} we present the
results of fits to a model function for the mass function $M$ and
extract the dynamically generated infrared quark mass; and in
Sec.~\ref{Sec:finite_vol} we discuss the possibility of finite volume
effects on $Z(p^2)$.  Finally, in Sec.~\ref{Sec:conclusions} we
present our conclusions and suggestions for further work.

\section{Improved quark propagators}
\label{Sec:improved_quarks}

A systematic program of improvement \cite{Symanzik:1983dc} proceeds by
adding all possible higher-dimensional local operators
to the Lagrangian.  When applied to the fermionic part of the QCD
action, adding all possible gauge invariant local dimension-5
operators yields the following Lagrangian
\cite{Luscher:1996sc,Dawson:1997gp}:
\begin{eqnarray}
{\mathcal L}(x) & = & {\mathcal L}^W 
- \frac{i}{4}c_{sw}a\psibar\sigma_{\mu\nu}F_{\mu\nu}\psi 
 + \frac{b_g am}{2g_0^2}\Tr(F_{\mu\nu}F_{\mu\nu})
 - b_mam^2\psibar\psi \nonumber \\
& & + c_1a\psibar\Dslash^2\psi
 + c_2am\psibar\Dslash\psi \; .
\label{eq:improve-action}
\end{eqnarray}
Here for notational brevity we introduce the simple notation $m$ for the
lattice bare mass, i.e., $m\equiv m^0(a)$.
In this equation we have used ${\mathcal L}^W$ for the standard Wilson
Lagrangian density and $(i/4)c_{sw}a\psibar\sigma_{\mu\nu}F_{\mu\nu}\psi$
is the so-called ``clover'' improvement term.  The sum of these two terms
is often referred to as the Sheikholeslami--Wohlert (SW) action.
That the action given by Eq.~(\ref{eq:improve-action}) is sufficient
to remove all $\order(a)$ errors has only been rigorously demonstrated
for on-shell quantities.  For gauge
dependent quantities, it is an open question whether further, gauge
non-invariant (but BRST invariant) terms must be added.  
We will assume that any such terms will be small.  We shall follow
the procedure used in studies of the gluon 
propagator\cite{Bonnet:2000kw,Leinweber:1999uu}, where it was seen
that a combination of improved actions and tree-level correction gave
reliable outcomes even at medium to high momenta.

Since the Wilson action explicitly breaks chiral symmetry,
the lattice bare mass should be taken as the so-called
subtracted bare mass
\cite{Luscher:1996sc}
\begin{equation}
 m \equiv m^0(a) \equiv m_0 - m_c
 = \frac{1}{a}\left(\frac{1}{2\kappa}-\frac{1}{2\kappa_c}\right) \;,
\label{eq:improve-field}
\end{equation}
where $m_0\equiv (1/2\kappa a)-4/a$ is the bare quark mass
appearing in the Wilson action.  At tree level, where
interactions are absent, the quark condensate will vanish when
the bare mass appearing in the action vanishes, i.e., when $m_0=0$ or
equivalently when $\kappa=1/8$.  In the interacting theory,
$\kappa_c$ is defined as the value of $\kappa$ at which the pion
mass vanishes and $m_c\equiv 1/(2\kappa_c a) - 4/a$ is a
nonperturbative fine-tuning correction needed to ensure that the
bare mass $m$ vanishes when the pion mass vanishes. 
The $b_g$ and $b_m$ terms correspond
to a (mass-dependent) rescaling of the coupling constant and the mass
respectively.  Since we we will work in the quenched approximation
we can set $b_g=0$.  The parameter $b_m$ will be absorbed into a
redefinition of the bare mass $m$ here and we will comment on this later.
At tree level, the  $c_1$ and $c_2$ terms can be
eliminated by the following transformation of the fermion field
\cite{Heatlie:1991kg},
\begin{eqnarray}
\psi & \to \psi' = &
 (1+b_q am)(1-c_qa\Dslash)\psi  \nonumber \\
\psibar & \to \psibar' = &
(1+b_q am)\psibar(1+c_qa\Dslashbak)
\label{eq:rotate}
\end{eqnarray}
In general, beyond tree level,
the improvement in the action must be combined with a
corresponding improvement in the fermion field \cite{Dawson:1997gp},
\begin{equation}
\psi' = (1+b'_q am)(1-c'_qa\Dslash)\psi + c_n\dslash\psi,
\label{eq:improved-psi}
\end{equation}
where the gauge dependent coefficient $c_n$ is needed when we compute
gauge dependent quantities, like the quark propagator.  By choosing
the correct improvement coefficients for the field, the $c_1$ and
$c_2$ terms may again be eliminated.
We note in passing that the coefficients $b'_q$ and $c'_q$ were
recently calculated at
one-loop level \cite{Capitani:2000xi}, while $c_n$ is still unknown.
However, here we will be restricting ourselves to tree-level
$\order(a)$-improvement throughout for the coefficients
$b'_q$, $c'_q$, and $c'_n$.  In that case, the $\order(a)$
improved action and fields after the transformation in
Eq.~(\ref{eq:rotate}) have
$b'_q\to b_q=\frac{1}{4}$ and
$c'_q\to c_q=\frac{1}{4}$, and $c_n=0$.
We will use the tree-level mean-field improved value for $c_{sw}$
and the nonperturbatively determined value of $\kappa_c$ to determine
the lattice bare mass $m$ in terms of $\kappa$.  Although we will
use tree-level improvement formulae for our quark actions and
propagators, it is more appropriate to use $m$ than $m_0$ in these.
Although there is an apparent inconsistency in using tree-level values
for $b'_q$ and $c'_q$ and the mean-field improved value for $c_{sw}$,
we have numerically verified that using mean-field improved values
for $b'_q$ and $c'_q$ makes no significant difference in practice.

The tree-level $\order(a)$-improved propagator can then 
be defined as
\begin{equation} 
S(x,y) \equiv \bra\psi'(x)\psibar'(y)\ket 
 = \bra (1+b_q am)^2(1-c_q a\Dslash(x))S_0(x,y;U)(1+c_q a\Dslashbak(y)) 
     \ket \; ,
\label{eq:rotprop}
\end{equation}
where $b_q=c_q=1/4$ and where
$S_0(x,y;U)$ for a given configuration
$U$ is simply defined as the inverse of the fermion matrix, 
\begin{equation}
M(x;U) \equiv
\Dslash_W(x;U)+(ia/4)c_{sw}\sigma_{\mu\nu}F_{\mu\nu}(x)+m
= \Dslash(x;U)+m+\order(a) \, ,
\end{equation}
where $\Dslash_W(x;U)$ is the lattice Wilson--Dirac
operator and $\Dslash(x;U)$ is the usual continuum covariant
derivative.  Therefore, $S_0(x,y;U)$ will always
satisfy the relations
\begin{eqnarray} 
 \bigl[\Dslash(x;U)+m\bigr]S_0(x,y;U) & = & \delta(x-y) + \order(a)
 \nonumber \\
 S_0(x,y;U)\bigl[-\Dslashbak(y;U)+m\bigr] & = & \delta(x-y) +
 \order(a)
\label{eq:eom}
\end{eqnarray}
The ``unimproved'' quark propagator $S_0$ will be defined here to be
that arising from the SW action consisting of the Wilson term and the
clover term, but with no other corrections.  Hence, $S_0$
is then given by the ensemble average of $S_0(x,y;U)$
\begin{equation} 
S_0(x,y) \equiv \bra S_0(x,y;U) \ket \; .
\label{eq:S_0}
\end{equation}
We will denote the tree-level $\order(a)$-improved
quark propagator obtained from Eq.~(\ref{eq:rotprop})
as the improved ``rotated'' propagator $S_R(x,y)$,
which is
\begin{equation}
S_R(x,y) \equiv \bra S_R(x,y;U)\ket
\equiv \bra (1+\frac{am}{2})\left[1-\frac{a}{4}\Dslash(x)\right]
S_0(x,y;U) \left[1+\frac{a}{4}\Dslashbak(y)\right] \ket \; .
\end{equation}

We can use Eq.~(\ref{eq:eom}) to obtain another, simpler expression
for the improved propagator from Eq.~(\ref{eq:rotprop})
\begin{equation}
\begin{split}
S(x,y) & = \bigl(1+2(b_q+c_q)am\bigr)S_0(x,y) - 2ac_q\delta(x-y) +
\order(a^2) \\
& = (1+am)S_0(x-y) - \frac{a}{2}\delta(x-y) + \order(a^2)\; ,
\end{split}
\label{eq:rotate-equiv}
\end{equation}
where we have used the fact that $b_q=c_q=1/4$
here.  We define the corresponding
version of the tree-level $\order(a)$-improved propagator as
\begin{equation}
S_I(x-y) \equiv (1+am)S_0(x-y) - \frac{a}{2}\delta(x-y)
\label{eq:improved-prop} 
\end{equation}
If we are only interested in on-shell improvement, e.g., hadronic
matrix elements, the $\delta$-function can be ignored.  However, it
is essential if we are considering off-shell properties such as the
quark propagator in momentum space.

In summary, we see that both $S_R$ and $S_I$ are tree-level improved
definitions of the SW--clover (i.e., the `Wilson plus clover')
propagator $S_0$.  However $S_R$ and $S_I$
will have different $\order(a^2)$ errors in general.
This will become an important
consideration when we later attempt to minimize lattice artifacts.

\section{Momentum space propagator}
\label{Sec:mom_space_quarks}

The momentum space quark propagator is given by
\begin{equation}
S(p) = \sum_x e^{-ipx}S(x,0)
\end{equation}
As is appropriate for fermions
we will be using periodic boundary conditions in the spatial
directions and antiperiodic boundary conditions in the time direction.
Hence, the available momentum values for an $N_i^3\times N_t$
lattice (with $N_i, N_t$ even numbers and $i=x,y,z$) are
\begin{alignat}{2}
p_i & = \frac{2\pi}{N_i a}\left(n_i - \frac{N_i}{2}\right)
 & ;\qquad
 n_i = & 1,2,\cdots,N_i \\
p_t & = \frac{2\pi}{N_t a} \left(n_t-\half-\frac{N_t}{2}\right)
& ;\qquad 
 n_t = & 1,2,\cdots,N_t \;.
\label{eq:latt_momenta}
\end{alignat}
We will also for notational convenience define the following `lattice
momenta': 
\begin{align}
k_\mu & \equiv  \frac{1}{a}\sin(p_\mu a) \\
\khat_\mu &
 \equiv  \frac{2}{a}\sin(p_{\mu}a/2) = \frac{\sqrt{2}}{a}\sqrt{1-\cos(p_\mu a)} 
\label{eq:latticemom} \\
\intertext{which differ by}
a^2\Delta k^2 & \equiv  \khat^2 - k^2 
 = \frac{a^2}{4}\sum_{\mu}p_\mu^4 + \order(a^4)
\end{align}

In the continuum, the quark propagator has the general from given by
Eq.~(\ref{Eq:generic-prop}).  On the lattice it is convenient
to work with the dimensionless quark propagator
$S(p)\equiv S^{\rm bare}(a;p)/a$.
We expect the lattice bare quark propagator to have a
similar form to its continuum equivalent, but with the
lattice momentum $\kslash$ replacing
$\pslash$, which can be appreciated by referring to the tree
level lattice propagators to be given later.
Because of hypercubic lattice artifacts
$Z^L$ and the dimensionless $M^L$ will be functions of $p_\mu$ rather
than $p^2$.  Hence we have for the dimensionless lattice bare quark
propagator the form
\begin{equation}
S(p) = \frac{1}{ia\kslash A(p) + B(p)} 
 \equiv \frac{Z^L(p)}{ia\kslash + M^L(p)}
 \equiv Z^L(p)\frac{-ia\kslash+M^L(p)}{a^2k^2+(M^L)^2(p)}
\label{Eq:AB-def}
\end{equation}
In the limit $a\to 0$ the continuum form will be recovered.
The (dimensionless) lattice functions $A(p)$ and $B(p)$ can
then easily be extracted from the inverse dimensionless
lattice quark propagator,
\begin{eqnarray}
A(p) \equiv \frac{1}{Z^L(p)} &
 = & \frac{-i}{4N_c k^2a^2}\Tr\left(\kslash S^{-1}(p)\right) \\
B(p) \equiv \frac{M^L(p)}{Z^L(p)}
 & = & \frac{1}{4N_c}\Tr\left(S^{-1}(p)\right)
\end{eqnarray}
In practice, however, it is easier to extract these functions without
inverting the propagator.  It is easily verified that
\begin{eqnarray}
A(p) & = & \frac{{\cal A}(p)}{k^2a^2{\cal A}^2(p)+{\cal B}^2(p)} \\
B(p) & = & \frac{{\cal B}(p)}{k^2a^2{\cal A}^2(p)+{\cal B}^2(p)}
\end{eqnarray}
where we have defined
\begin{equation}
{\cal A}(p) \equiv \frac{i}{4N_c k^2a^2}\Tr\left(\kslash S(p)\right) \qquad 
{\cal B}(p) \equiv \frac{1}{4N_c}\Tr S(p)
\end{equation}

\subsection{Tree-level expressions}
\label{Sec:tree_level_exp}

As was done in the introduction in Sec.~\ref{Sec:intro} we will use
the superscript $(0)$ to denote the tree-level versions of each of the
quark propagator definitions and actions considered.  These are simply
the propagators that would be obtained from the various definitions
when the interaction with the gluons is turned off, i.e., when all the
gluon links are taken to be the identity.  We will also be writing $m$
rather than $m_0$ throughout, since at tree level the two are
identical.  The dimensionless SW fermion propagator at tree level is
identical to the pure Wilson propagator and is given by
\cite{Karsten:1981wd,Carpenter:1985dd}
\begin{equation}
S_0^{(0)}(p) = \frac{-i\kslash a + ma + \half\khat^2 a^2}
{k^2 a^2 + \left(ma+\half\khat^2 a^2\right)^2} \, .
\end{equation}
The tree-level form of the tree-level $\order(a)$-improved
propagator $S_I$ is given by
\begin{equation}
S_I^{(0)}(p) = (1+ma)S_0^{(0)}(p)-\half \, .
\end{equation}
If we write
\begin{equation}
\left(S_I^{(0)}(p)\right)^{-1} = i\kslash a A_I^{(0)}(p) + B_I^{(0)}(p) \, ,
\end{equation}
we find
\begin{eqnarray}
A_I^{(0)}(p) & = &
 -\frac{i}{4N_c}\Tr\left[\kslash a\left(S_I^{(0)}(p)\right)^{-1}\right]/k^2 a^2
  = 1 + \order(a^2) \\
B_I^{(0)}(p) & = & \frac{1}{4N_c}\Tr\left(S_I^{(0)}(p)\right)^{-1} 
 = ma\left(1-\frac{ma}{2} + \order(a^2) \right) 
\end{eqnarray}
We see that the quark mass gets an $\order(a)$ correction.  The purpose
of the improvement term $b_m$ in the action 
(\ref{eq:improve-action}) was to cancel this change in the bare mass
$m$.   However, by omitting this correction we have
have simply absorbed it into a redefinition
of $m$.  $A^{(0)}(p)$ is equal to unity up to $\order(a^2)$, as
expected.  The details of the derivation are given in the appendix.

\begin{figure}
\begin{center}
\setlength{\unitlength}{1cm}
\setlength{\fboxsep}{0cm}
\begin{picture}(14,14)
\put(0,7){\begin{picture}(7,7)\put(-0.9,-0.4){\ebox{8cm}{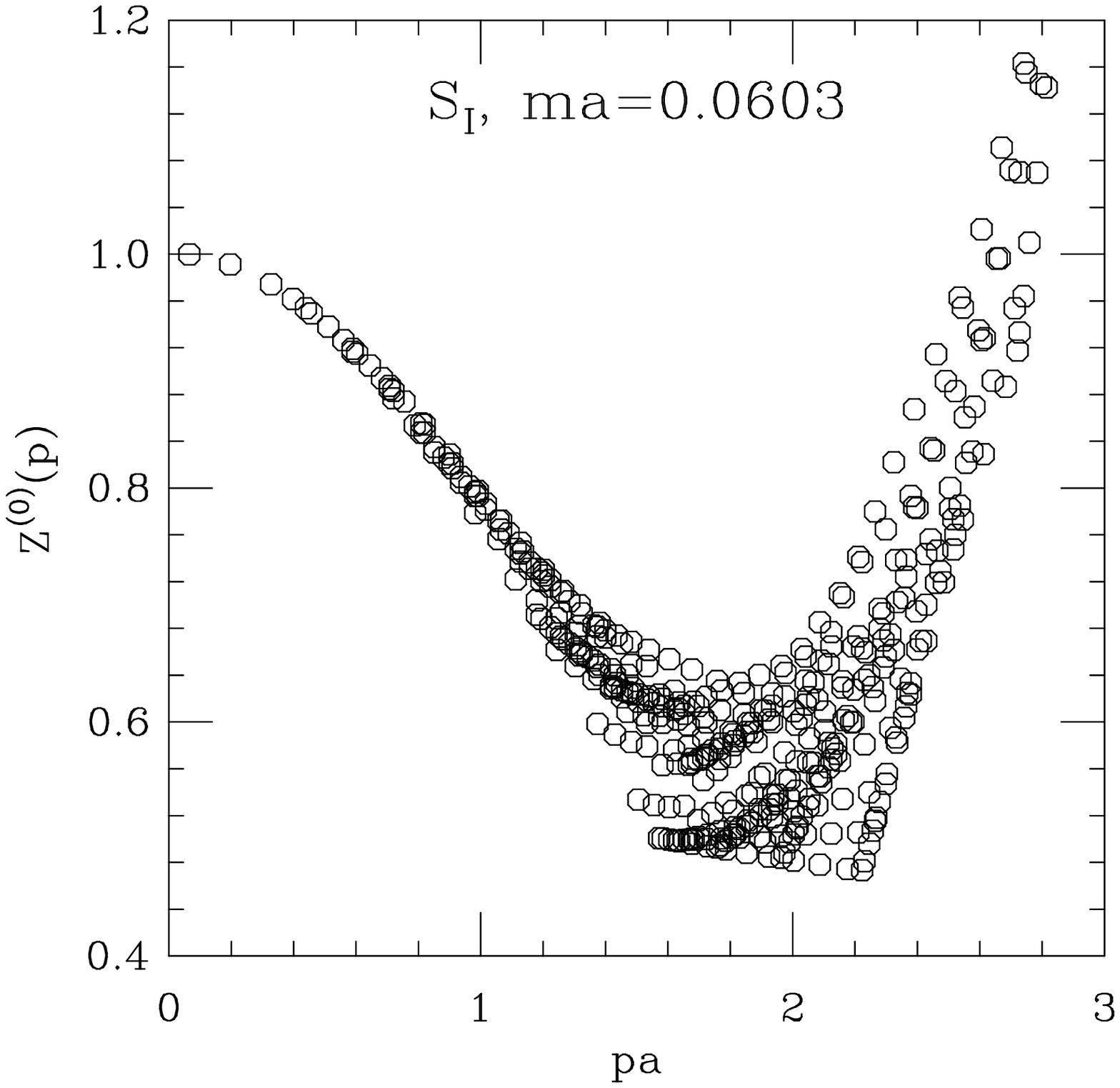}}\end{picture}}
\put(7,7){\begin{picture}(7,7)\put(-0.9,-0.4){\ebox{8cm}{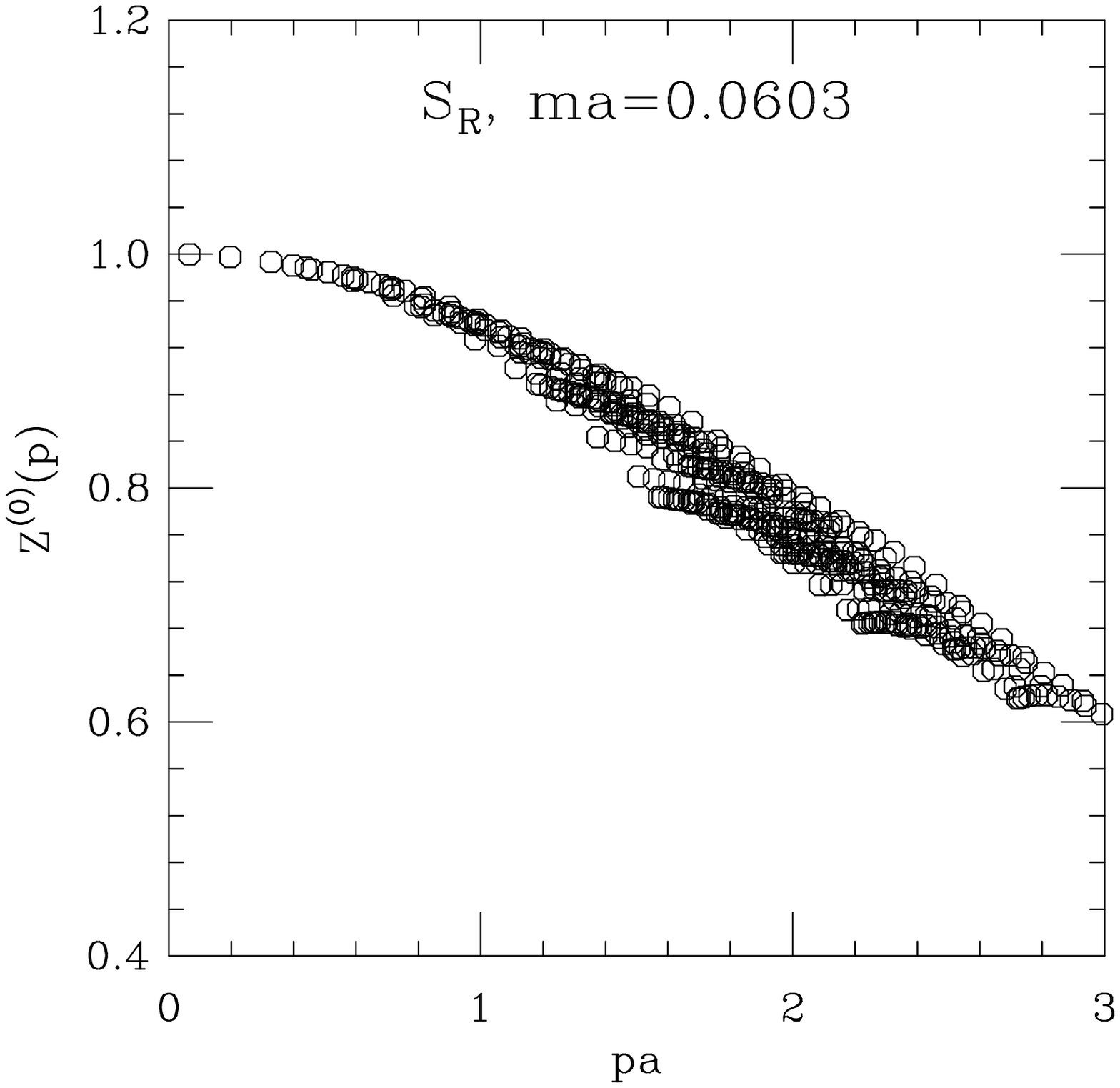}}\end{picture}}
\put(0,0){\begin{picture}(7,7)\put(-0.9,-0.4){\ebox{8cm}{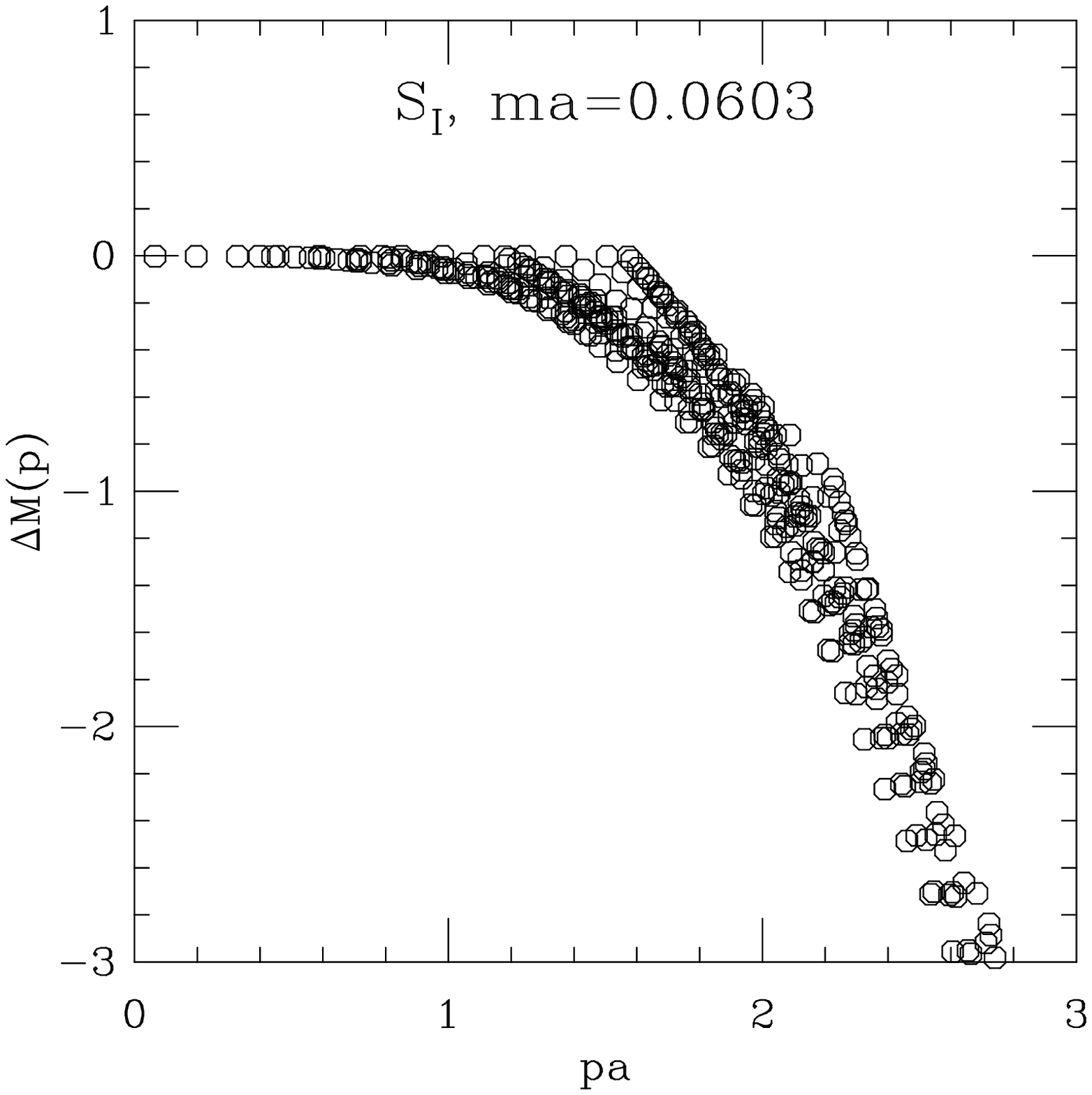}}\end{picture}}
\put(7,0){\begin{picture}(7,7)\put(-0.9,-0.4){\ebox{8cm}{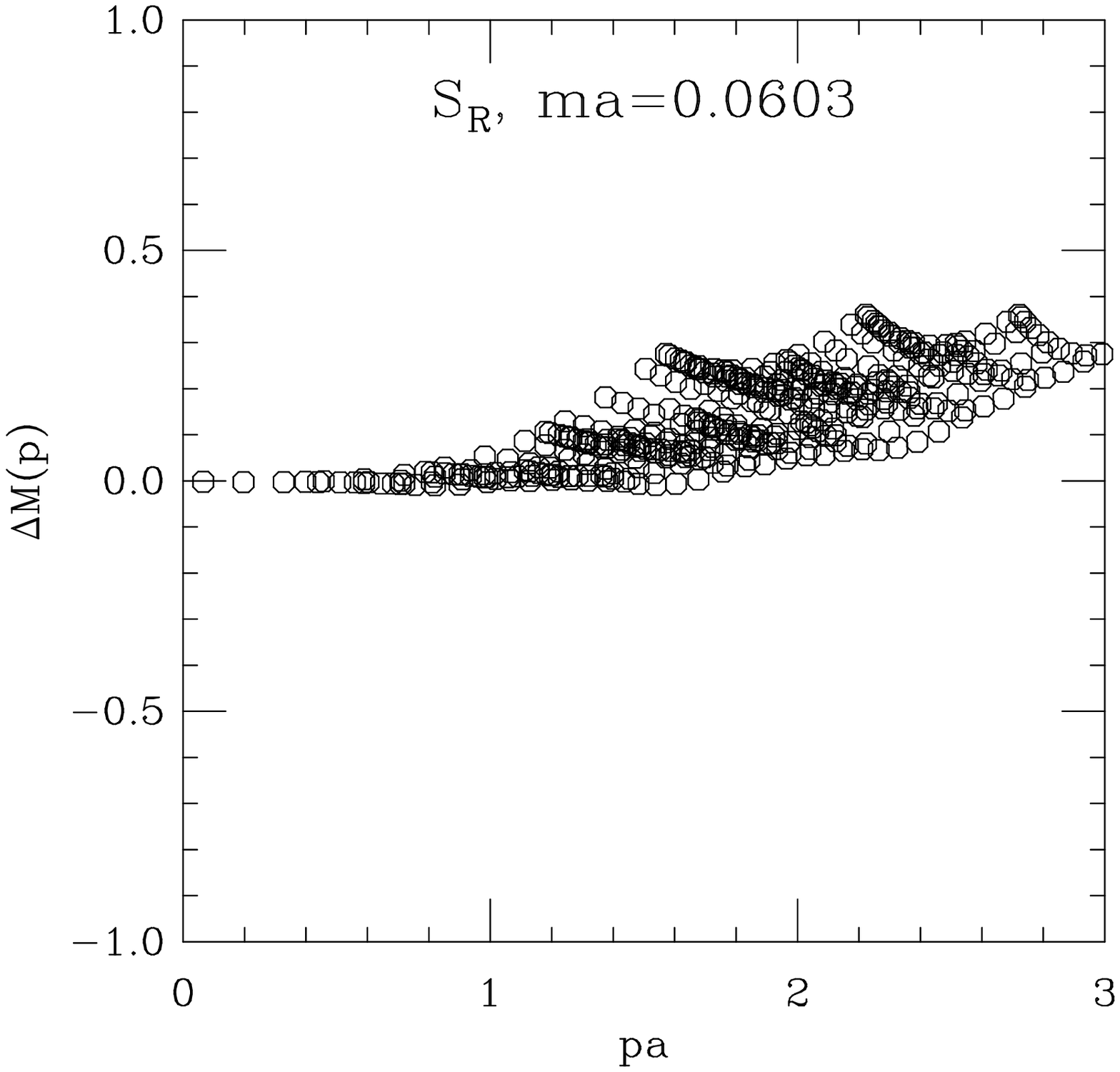}}\end{picture}}
\end{picture}
\end{center}
\caption{ Plots of the analytic functions
$\zz$ (top) and $a\dmz$ (bottom) versus the momentum
$pa$ for $S_I(p)$ (left) and $S_R(p)$ (right).  The results shown here
are obtained from the analytical expressions in
Eqs.~(\protect\ref{Eq:z0-imp})--(\protect\ref{b0-rot}).}
\label{Fig:z0m0}
\end{figure}

It is also useful to write the propagator in the following way,
\begin{equation}
\left(S^{(0)}(p)\right)^{-1} = \frac{1}{\zz(p)}
 \left[ i\kslash a + ma + a\dmz(p)\right] \, .
\end{equation}
The analytic expressions for $\zz(p)$ and 
$\dmz(p)$ for both the improved
propagators $S_I$ and $S_R$ are given in the appendix.
To illustrate the behavior of these tree-level functions we show in
Fig.~\ref{Fig:z0m0} the forms of $\zz$ and $\dmz$ for both
of our improved actions $S_I$ and $S_R$.  The horizontal axis
is $pa\equiv \sqrt{p_t^2+p_x^2+p_y^2+p_z^2}$, where the possible values
of the momenta are those given in Eq.~(\ref{eq:latt_momenta}).
Since $\zz(p)$ deviates from 1 and $\dmz(p)$ deviates
from zero at medium to high momenta, it is immediately obvious
that the finite $a$ effects are very large
and we will need some method for taking care of them if we are to
obtain physically meaningful results.  The tree-level behavior is
particularly pathological for $S_I(p)$, with finite-$a$ effects
of several hundred percent appearing in $\zz$, and
$\dmz$ being many times larger than $m$, and negative.
The spread in the points is due to hypercubic artifacts, since
on the lattice $\zz$ and $\dmz$ are functions of
$p_\mu$ and not $p^2$.  The finite-$a$ effects in $S_R$ are much more
mild and offer the hope that they might be partially compensated for.
Clearly in the limit $a\to 0$ we recover the continuum result
where $\zz(p)=1$ and $\dmz(p)=0$ for all $p$.

\section{Analysis}
\label{Sec:analysis}

\subsection{Tree-level correction}
\label{Sec:tree_level_sub}

Recall that the quark propagators calculated on the lattice are
actually the bare quark propagators, which become the tree-level
propagators when the interactions are switched off.
We know that QCD is asymptotically free, which means that at sufficiently
high momentum values the bare quark propagator should
approach the tree-level quark propagator up to logarithmic
corrections, i.e., on the lattice for large momenta we should find
$S(p)\to S^{(0)}(p)$ up to logarithms.  The deviation of these
from each other is a direct measure of the nonperturbative
effects due to the interactions felt by the quarks.  Hence,
we are here primarily interested in studying the {\em deviation} of the
quark propagator from its tree-level form.

We will attempt to
separate out the tree-level behavior by writing
\begin{equation}
  S^{-1}(pa) = \frac{1}{Z(pa)\zz(pa)} 
  \left[ia\kslash + aM(pa) + a\dmz(pa) \right] .
\label{eq:zmdef}
\end{equation}
If asymptotic freedom holds for the momentum range we are considering,
we should expect that $Z(pa)\to 1$ and $M(pa)\to m$ (up to
logarithmic corrections) for large $p$.

Equation (\ref{eq:zmdef}) can be rewritten to yield expressions for
$Z(pa)$ and $M(pa)$ in terms of the functions $Z^L$ and $M^L$ (or,
equivalently, $A$ and $B$) defined in
Eq.~(\ref{Eq:AB-def}), 
\begin{eqnarray}
Z(pa) & = & \frac{Z^L(pa)}{\zz(pa)} \\
aM(pa) & = & M^L(pa) - a\dmz(pa)
\end{eqnarray} 
We refer to the functions $Z$ and $M$ obtained in this way as the
{\em tree-level corrected} forms of the lattice quantities
$Z^L$ and $M^L$.

It is important that one not be confused by the different uses of the
expression ``tree-level''.  Firstly, there is an $\order(a)$-improvement
which was only implemented at tree-level rather than having the
improvement coefficients determined in some nonperturbative way.
Secondly there are tree-level propagators which are the bare
propagators when there are no interactions.  Thirdly, we have just
now introduced our method of tree-level correction, which will
hopefully minimize the finite-$a$ errors in our extraction of the
quark propagator from the lattice.
Because the tree-level behavior of $S_I$, i.e., $S^{(0)}_I$,
is much worse at medium and high momenta than that of $S_R$,
we anticipate that our tree-level correction
may not be adequate in that case.
We therefore expect the tree-level correction
method to work significantly better for $S_R$ than for $S_I$.

\subsection{Cuts}
\label{sec:cuts}

Even after the tree-level correction has been performed, there will
still be anisotropies in the data, resulting from finite $a$ effects
beyond tree level.  To remove these, we select momenta lying close
to the diagonal in momentum space.
We choose the diagonal because finite-$a$ hypercubic artifacts
will be minimized at a given $pa$ when the momentum is
approximately equally spread among the four momentum components.
Ideally, one should attempt an $a\to 0$ extrapolation, but given the
available data we will see that this cut on the data removes
most of these artifacts.
We define the distance of a point from the diagonal by
\begin{equation}
\Delta p = |p|\sin\theta(p) \, ,
\label{eq:deltaq-def}
\end{equation} 
where the angle $\theta(p)$ is given by
\begin{equation}
\cos\theta(p) = \frac{p\cdot\hat{n}}{|p|} \, ,
\label{eq:angle-def}
\end{equation}
and $\hat{n} = \half(1,1,1,1)$ is the unit vector along the diagonal.
We select momentum values such that $\Delta p \leq \pi/8a$, ie.\ within
one unit of spatial momentum from the diagonal.  We refer to this
selection as the ``cylinder cut'', since the momenta selected lie within
a cylinder around the diagonal in momentum space.

\section{Results}
\label{Sec:results}

The quark propagator is calculated at $\beta=6.0$ on a $16^3\times 48$
lattice, using the tree-level mean-field (`tadpole') improved value
$c_{sw}=1.479$.  For this action with these parameter values,
$\kappa_c$ is found to be $\kappa_c=0.1392$ \cite{Bowler:1999ae}. Two
values for $\kappa$ were used: $\kappa=0.137$, corresponding to
$ma=0.0603$, and $\kappa=0.1381$, corresponding to $ma=0.031$.  
The lattice spacing as determined from string tension measurements
in the gluon sector at $\beta=6.0$ is
$a=0.106\pm 0.002$~fm  or equivalently 
$1/a = 1.855 \pm 38$~GeV and so the values
for the quark masses are $m=112$~MeV and $m=57.5$~MeV respectively.
 
The configurations were fixed to Landau gauge with an accuracy of
$\theta\equiv\sum_{x,\mu}|\partial_\mu A_\mu(x)|^2 < 10^{-12}$.  At
$\kappa=0.137$, we have generated both $S_0$ and $S_R$.  At
$\kappa=0.1381$, only $S_0$ was generated.  $S_I$ is easily
constructed from $S_0$.  We have used the tree-level values for the
coefficients $b_q$ and $c_q$ as previously stated, rather than the
mean-field improved
values, as the difference between the two is negligible compared to
the $\order(a^2)$ and higher effects which the tree-level correction
scheme attempts to minimize.  We have explicitly verified that
replacing the tree-level values for $b_q$ and $c_q$ with the
mean-field improved values makes only negligible difference.  However,
in any future study it would clearly be preferable to make consistent
use of mean-field improved or non-perturbatively determined (as far as
they are available) improvement coefficients throughout.
All the results shown for $S_R$ are for 20
configurations, while the results for $S_I$ are for 499
configurations, unless otherwise specified.

As a further check on our results, we have also analyzed 60
configurations at $\beta=5.7$ on a $12^3\times 24$ lattice, for
$\kappa=0.13843$ and $\kappa=0.14077$, corresponding to $ma=0.128$
and 0.068 respectively.  Here, $\kappa_c=0.1432$.  In this case, all
three propagators were 
generated for all configurations.  We will not explicitly
show these results here but will comment on their relevance in our
later discussion.

\subsection{Uncorrected data}
\label{Sec:unsubt_data}

\begin{figure}
\begin{center}
\setlength{\unitlength}{1cm}
\setlength{\fboxsep}{0cm}
\begin{picture}(14,7)
\put(0,0){\begin{picture}(7,7)\put(-0.9,-0.4){\ebox{8cm}{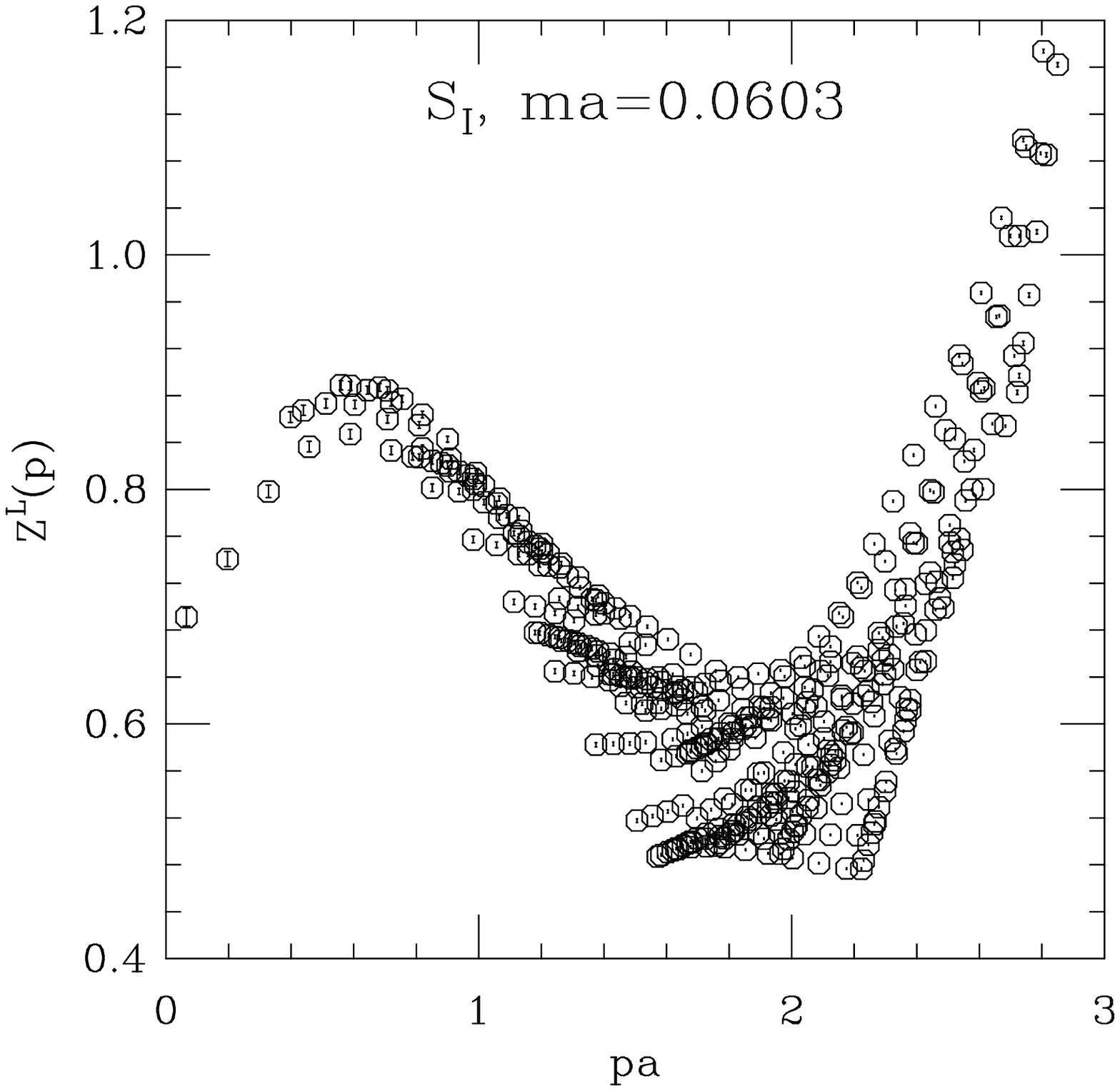}}\end{picture}}
\put(7,0){\begin{picture}(7,7)\put(-0.9,-0.4){\ebox{8cm}{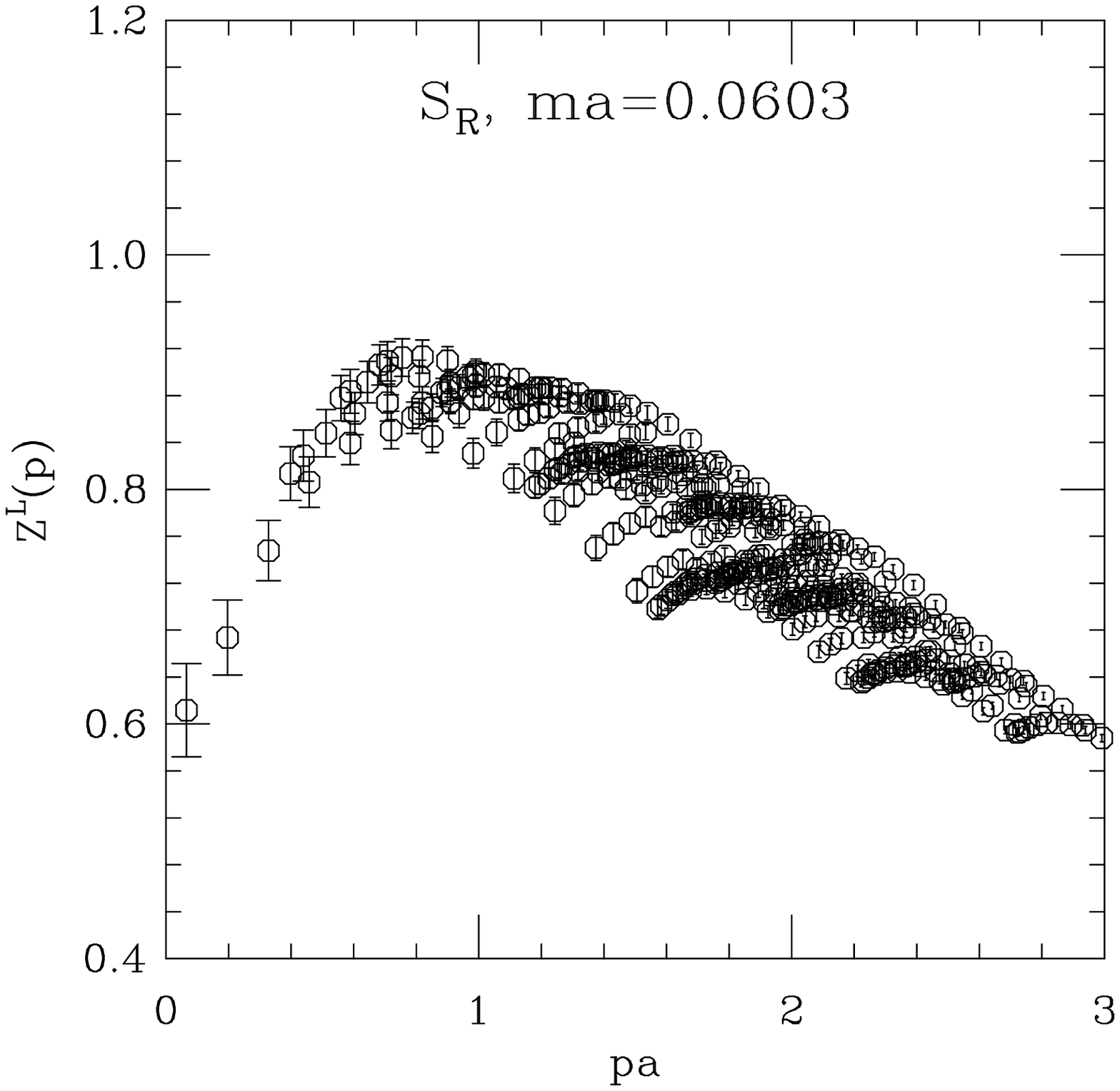}}\end{picture}}
\end{picture}
\end{center}
\caption{
$Z^L(p)=1/A(p)$  as a function of momentum $p$ for $S_I(p)$ (left) and for
$S_R(p)$ (right) with the bare quark mass corresponding to $\kappa=0.137$.
No tree-level correction has been made and no data cuts have been applied. 
}
\label{Fig:zrotcompare}
\end{figure}

\begin{figure}
\begin{center}
\setlength{\unitlength}{1cm}
\setlength{\fboxsep}{0cm}
\begin{picture}(14,7)
\put(0,0){\begin{picture}(7,7)\put(-0.9,-0.4){\ebox{8cm}{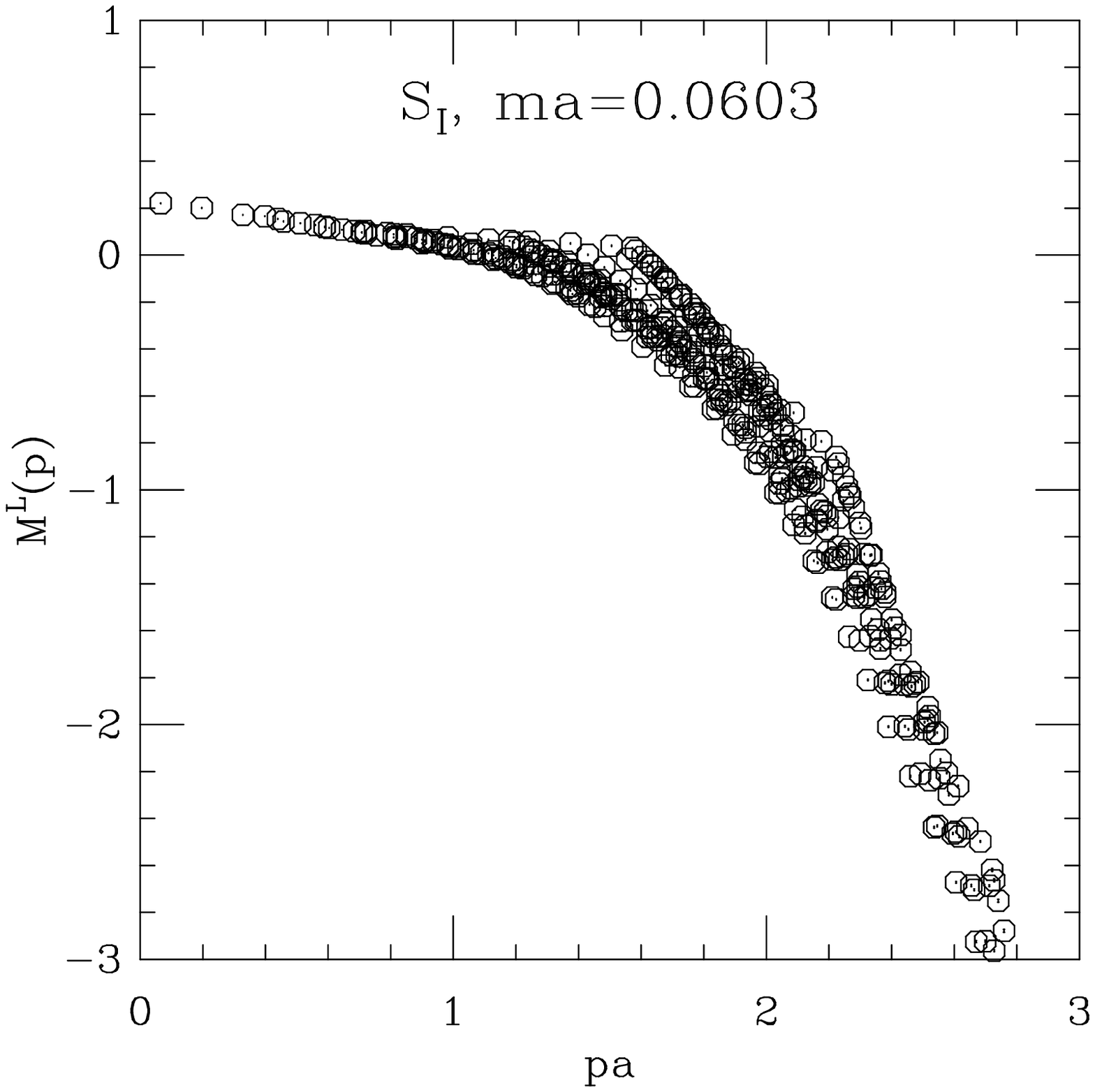}}\end{picture}}
\put(7,0){\begin{picture}(7,7)\put(-0.9,-0.4){\ebox{8cm}{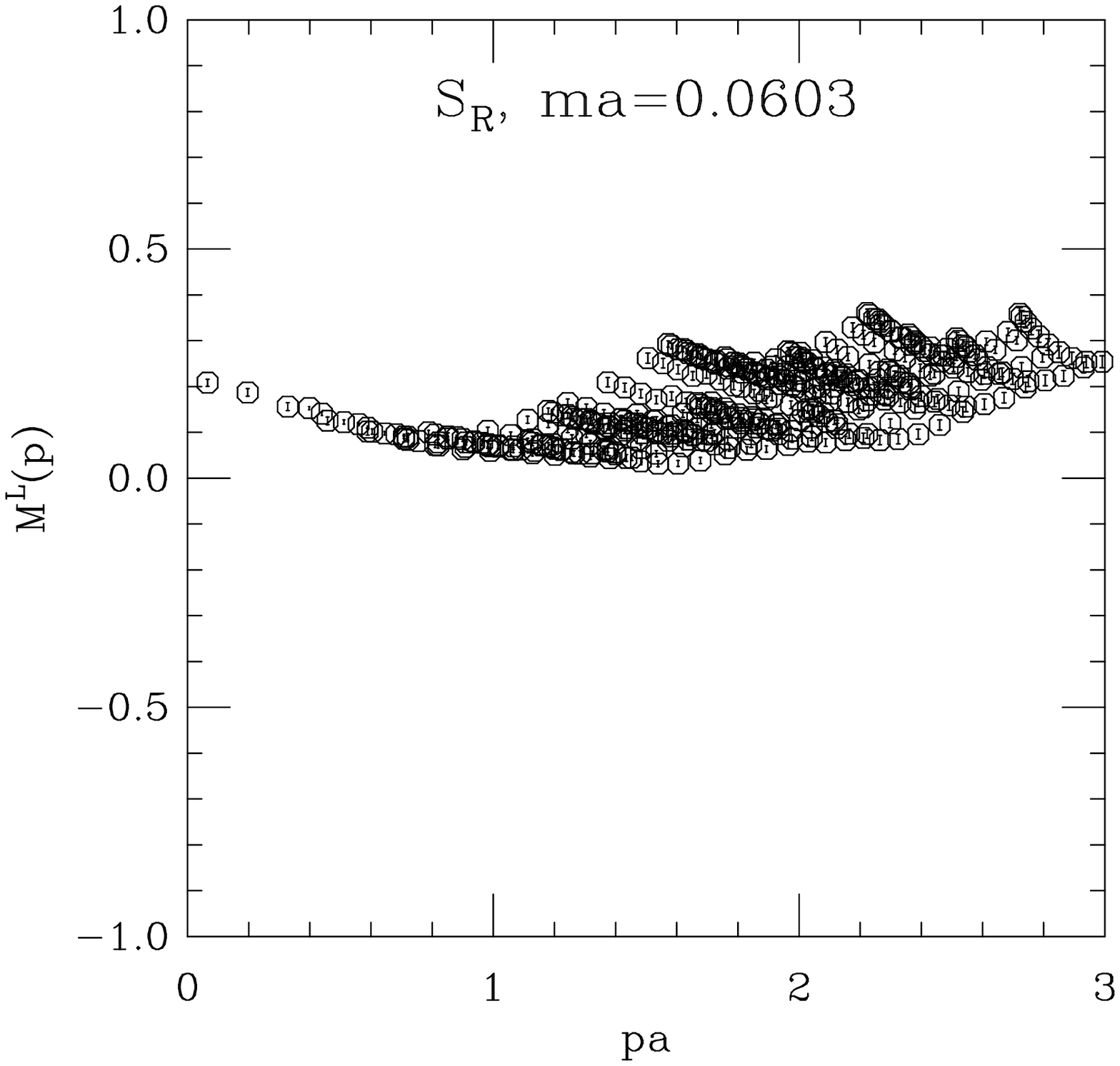}}\end{picture}}
\end{picture}
\end{center}
\caption{
$M^L(p)$ for $S_I(p)$ (left) and for
$S_R(p)$ (right) with the bare quark mass corresponding to 
$\kappa=0.137$.
No tree-level correction has been made and no data cuts have been applied. 
}
\label{Fig:mrotcompare}
\end{figure}

Let us first see what happens when we use the na\"{\i}ve formulae for
$Z$ and $M$ without implementing the tree-level correction, i.e.,
we first consider $Z^L$ and $M^L$.
Since Eq.~(\ref{eq:eom}) holds precisely configuration by configuration,
Eq.~(\ref{eq:rotate-equiv}) should be satisfied non-perturbatively.
However, the $\order(a^2)$-term can be quite large.
In Fig.~\ref{Fig:zrotcompare} we show $Z^L(p)=1/A(p)$ as a function of
$pa$ using $S_I(p)$ and $S_R(p)$ respectively, while in
Fig.~\ref{Fig:mrotcompare} we present $M^L(p)$ for
the same two cases.
Comparing these figures with the tree-level behavior shown in
Fig.~\ref{Fig:z0m0} we see that finite-$a$ errors completely dominate
these (uncorrected) quark propagators at medium and high momenta.
Only in the infrared, below $pa\lesssim 0.8$, might we be able to extract
physically significant information.

A comparison with the `unimproved' SW propagator $S_0$ shows that both
$S_I(p)$ and $S_R(p)$ are considerably better behaved in the infrared
than the `na\"{\i}ve' propagator $S_0(p)$.  In particular, the mass
function is a decreasing function of $pa$ up to $pa\sim1$, which is
what one would expect from asymptotic freedom.  This is not the case
for $S_0(p)$, which begins to increase monotonically in the infrared
at $pa\sim 0.4$ as does its tree-level form.
Also, the values for $Z^L(p)=1/A(p)$ and
$M^L(p)$ agree for $S_R$ and $S_I$ within
errors up to $pa\sim0.8$, while the values
obtained from $ S_0(p)$ are significantly different from the improved values
even at low momenta.

At momenta above $pa\sim1$ the $\order(p^2 a^2)$ and higher terms
dominate and it is impossible to extract any meaningful information
from these uncorrected data.  This can be appreciated most
dramatically by the way the lattice data diverge with increasing
momentum for the two improved (but uncorrected) propagators.

\subsection{Tree-level corrected data}
\label{Sec:tree_level_subtr_data}

Since, as we saw in the previous section, the tree-level form
completely dominates the high-momentum data, we may hope that by
factoring out this behavior we will get something which lies close to
the continuum asymptotic form.  In the low and intermediate momentum
region we may then be able to extract the physical, nonperturbative
behavior of the functions $Z(p)$ and $M(p)$.

\begin{figure}
\begin{center}
\setlength{\unitlength}{1cm}
\setlength{\fboxsep}{0cm}
\begin{picture}(14,7)
\put(0,0){\begin{picture}(7,7)\put(-0.9,-0.4){\ebox{8cm}{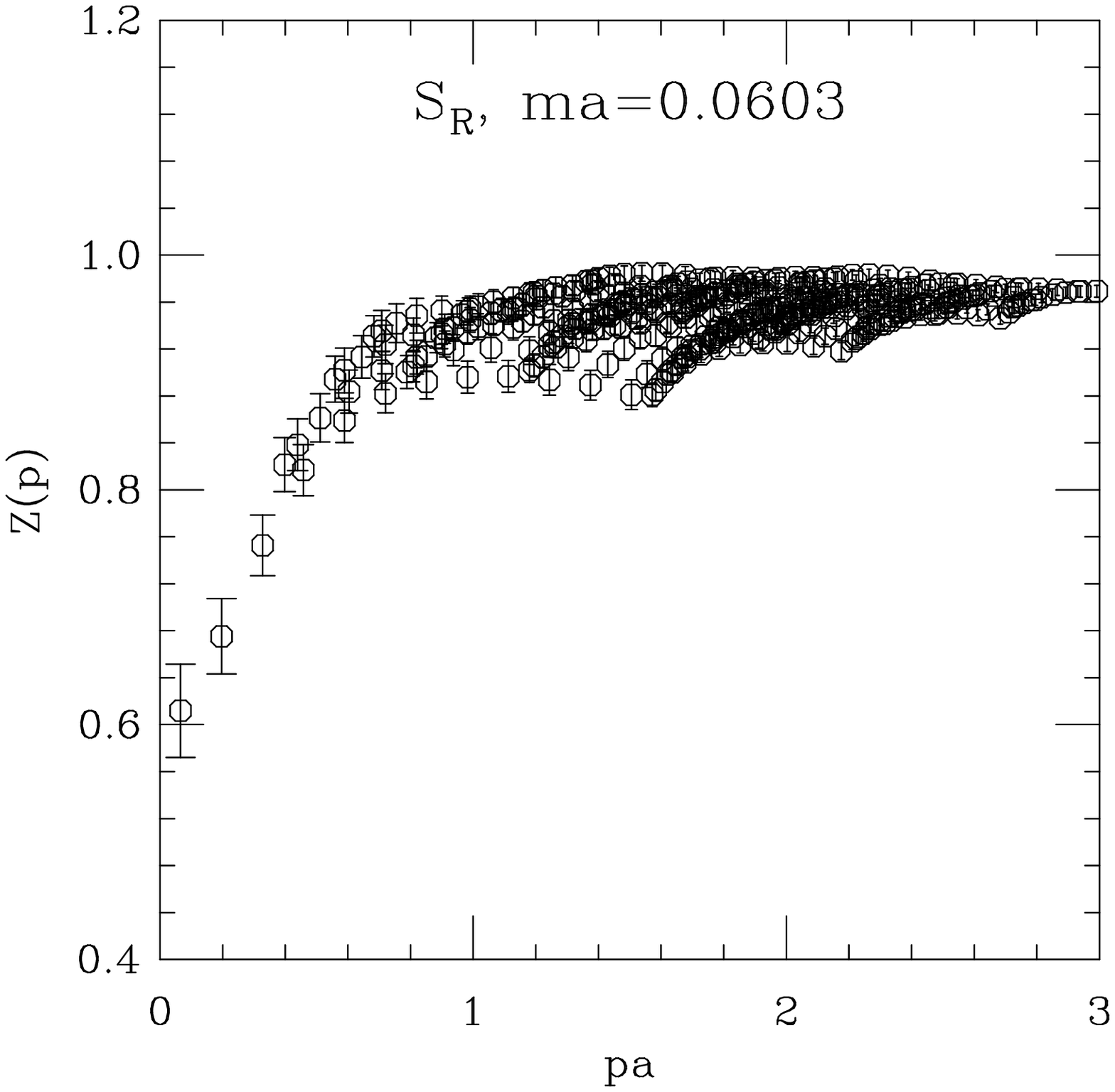}}\end{picture}}
\put(7,0){\begin{picture}(7,7)\put(-0.9,-0.4){\ebox{8cm}{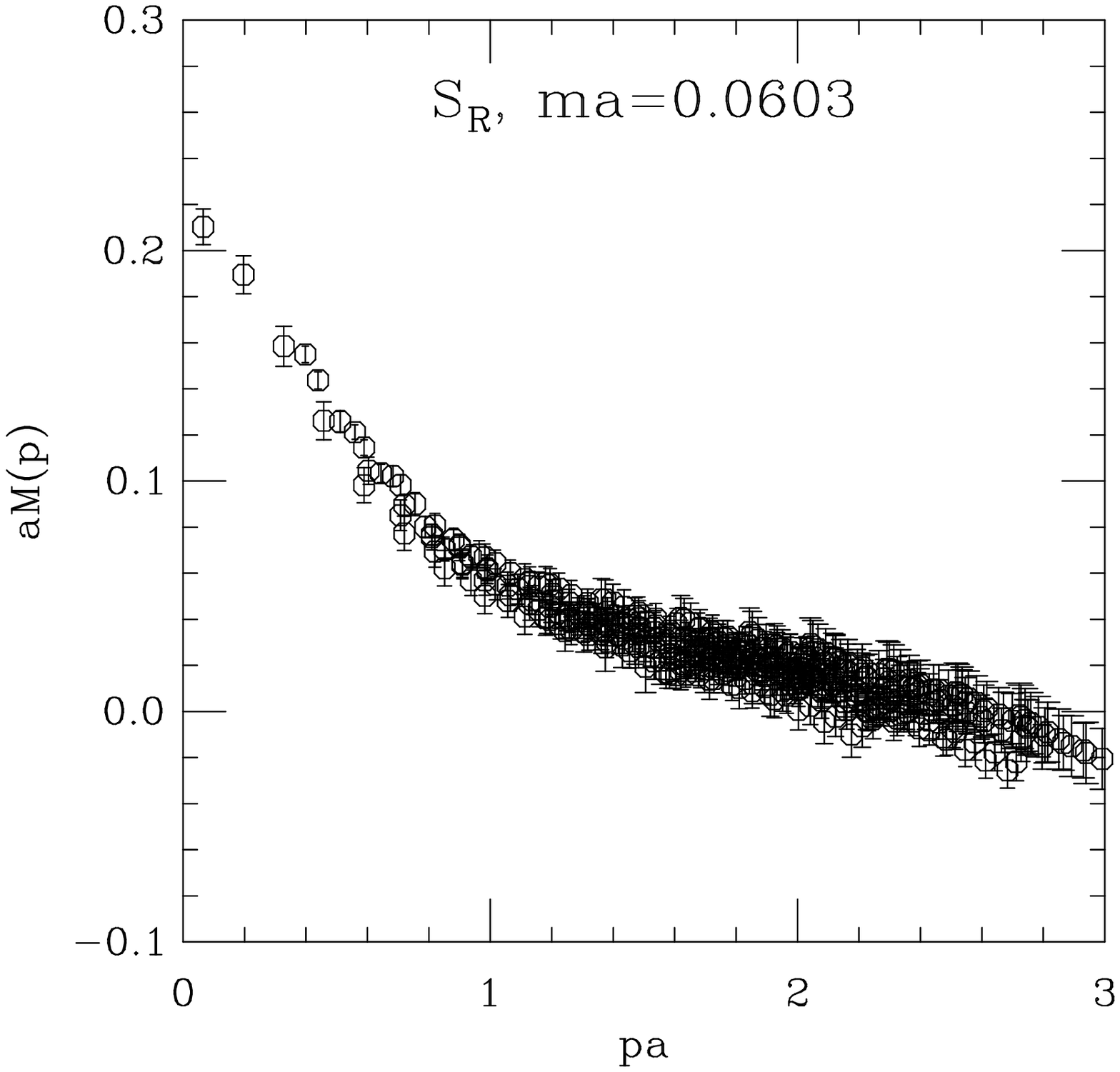}}\end{picture}}
\end{picture}
\end{center}
\caption{ 
$Z(p)$ (left) and $aM(p)$ (right) for our
preferred form of the improved propagator $S_R(p)$ at $\kappa=0.137$.
The lattice data shown are obtained using the tree-level correction
defined in Eq.~(\protect\ref{eq:zmdef}) but without any cuts.}
\label{Fig:zmsub}
\end{figure}

When applying this correction, we find that there is a dramatic
improvement in the behavior with $pa$ of all three of our
forms of the propagator, i.e., for $S_0, S_I$, and $S_R$.
However, the pathological
behavior of $S_I^{(0)}(p)$ at high momenta gives rise to a
cancellation of large terms in the subtracted mass, leading to a
behavior for the mass function which is clearly at odds with the
expectation from asymptotic freedom.  Thus, as expected, the finite-$a$
errors in $S_I$ are simply too large to be corrected by our simple
tree-level correction procedure.  As previously noted, our
unimproved propagator $S_0$ behaves poorly even at very low momenta and
cannot therefore be trusted.  It is therefore desirable to use
the definition $S_R$ for the improved propagator\footnote{It is
reassuring that a similar
conclusion was reached in Ref.~\cite{Cudell:1999kf}.}
and to apply our tree-level correction to that.  The results
for $Z(p)$ and $M(p)$ for our preferred propagator $S_R(p)$ are shown
in Fig.~\ref{Fig:zmsub} as functions of $pa$ for $\kappa=0.137$.
We see that the medium to large momentum behavior has been dramatically
improved by our tree-level correction procedure as expected, i.e., it
behaves in a way reasonably consistent with the expectations
of asymptotic freedom.  The spread of the lattice data due to hypercubic
artifacts is somewhat reduced but has not been eliminated.

\begin{figure}
\begin{center}
\setlength{\unitlength}{1cm}
\setlength{\fboxsep}{0cm}
\begin{picture}(14,7)
\put(0,0){\begin{picture}(7,7)\put(-0.9,-0.4){\ebox{8cm}{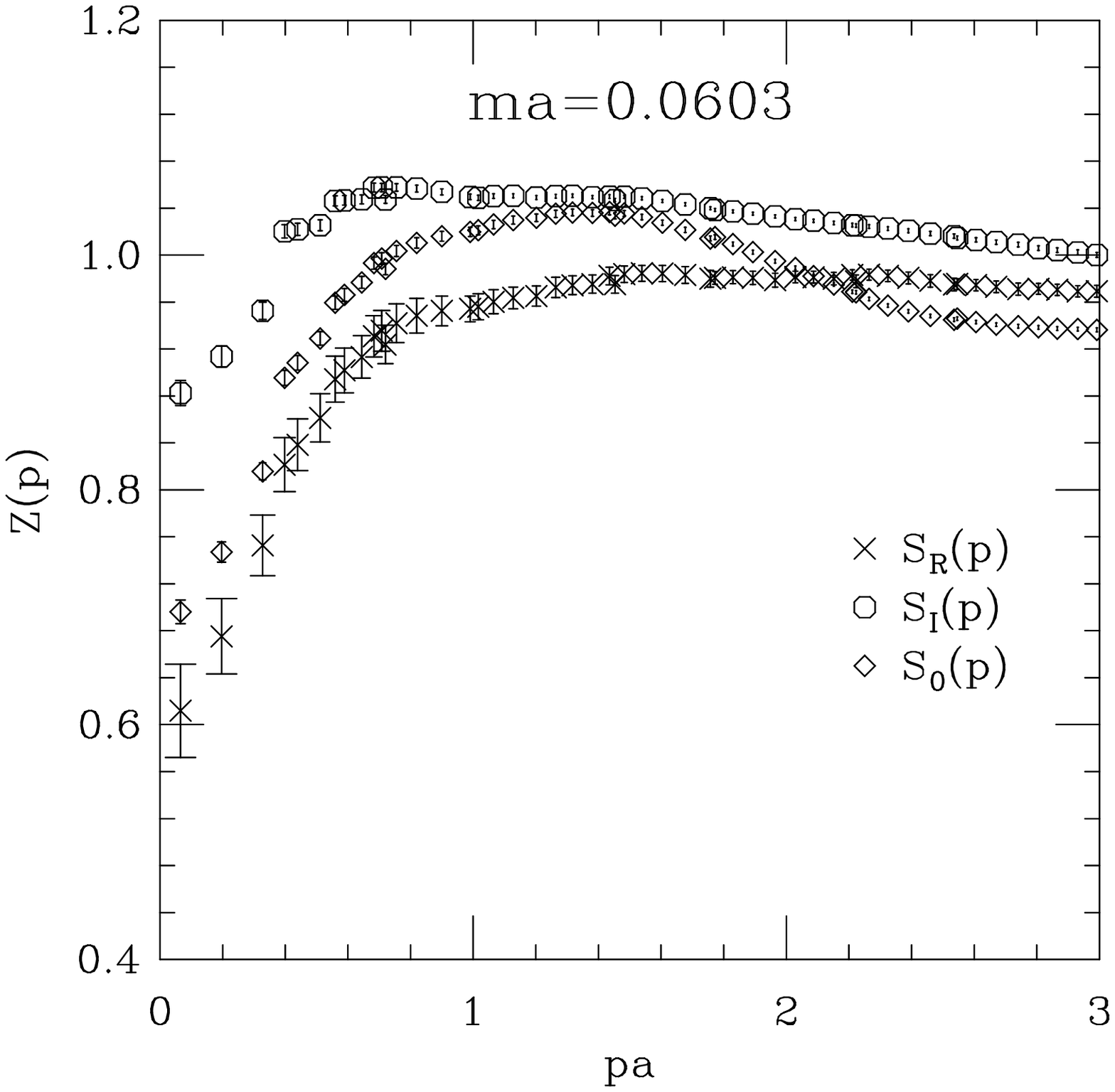}}\end{picture}}
\put(7,0){\begin{picture}(7,7)\put(-0.9,-0.4){\ebox{8cm}{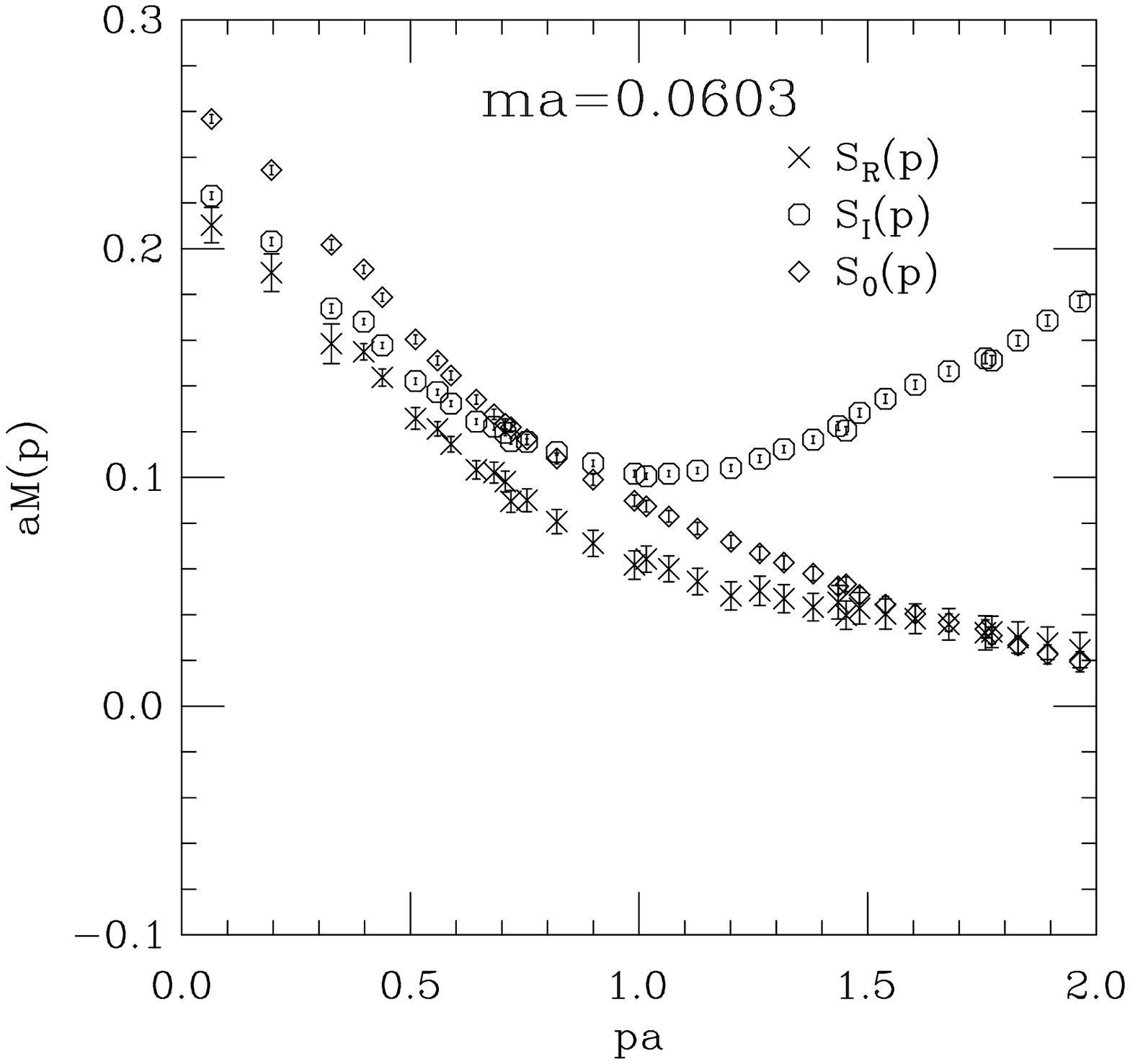}}\end{picture}}
\end{picture}
\end{center}
\caption{ The tree-level corrected functions $Z(p)$ (left) and $aM(p)$
(right) for all three propagators, after performing the cuts described
in section~\protect\ref{sec:cuts}.  The values for $S_R$ are obtained
from 20 configurations; for the two other propagators the data are
from 300 configurations.  Note that the
$Z(p)$ functions need to be scaled to agree at some ``safe'' momentum
scale (e.g., $\mu a\simeq 0.4$) before being compared.  }
\label{Fig:cutscompare}
\end{figure}

In Fig.~\ref{Fig:cutscompare} we show the lattice
results for $Z$ and $M$ for all three definitions of our quark
propagator, after implementing
both the tree-level correction and the cylinder cuts described in
section~\ref{sec:cuts}.
The tree-level correction is clearly failing for $S_I$ as is evident
from the behavior of the mass function $M(p)$.
Although the behavior of the unimproved propagator $S_0$ has been
considerably improved, as we previously observed it cannot be
trusted even at relatively low momenta and so it must be discarded.
The apparent difference in the
behavior of $Z$ between the two improved definitions of the
propagator $S_I$ and $S_R$, even at relatively low momenta, is at
first sight puzzling.  However, it must be recalled that
$Z(p)\equiv Z_2(\mu;a) Z(\mu;p)$ and that it is actually
$Z(\mu;p)$ that we should be comparing for the
different actions.  Different actions will in general have different
values of the renormalization constant $Z_2(\mu;a)$.  If we renormalize
at some ``safe'' momentum scale where we would expect both
improved propagators to be reliable, e.g., $\mu a\sim 0.4$, then
the apparent difference is much reduced except at medium to
high momenta where $S_I$ can no longer be trusted.
Analysis of the data at $\beta=5.7$, which corresponds to a coarser
lattice, and hence larger finite-$a$ effects, is also consistent with
this interpretation.
Below $pa\sim0.6$, the values for $M(p)$ agree within errors for the
two versions of the improved propagator.  In particular, the value for
the infrared mass $M(p\to0)$ agrees well.  In contrast, and
not surprisingly, the unimproved propagator $S_0$
yields a mass which is 3--4$\sigma$ higher.
We again clearly see from this plot the poor high-momentum behavior
of $M(p)$ arising from the inexact cancellation of large
finite-$a$ errors in $S_I$.

\begin{figure}
\begin{center}
\setlength{\unitlength}{1cm}
\setlength{\fboxsep}{0cm}
\begin{picture}(8,8)
\put(-0.5,-0.5){\ebox{9cm}{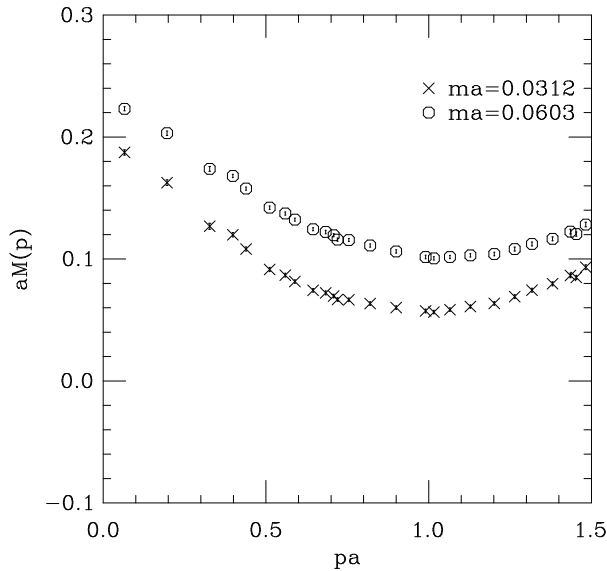}}\end{picture}
\end{center}
\caption{ The tree-level corrected mass function $M(p)$ from $S_I$,
for $\kappa=0.137$ (circles) and $\kappa=0.381$ (crosses), after
performing the cuts described in section~\protect\ref{sec:cuts}.  The
increase in $M(p)$ for $pa>1$ is an indication of the difficulty of
accurately subtracting off the tree-level mass function for this
definition of the improved propagator.  $S_I$ has been used
in this case because we lack data for more than one quark mass for our
preferred definition $S_R$; however, the mass functions agree well in the
infrared.  The data shown are from 300 configurations.  }
\label{Fig:masscompare}
\end{figure}

Fig.~\ref{Fig:masscompare} shows $M(p)$ calculated from $S_I$ for
the two values of the quark mass.  In the infrared region, the mass 
changes only slightly as the bare quark mass is halved, pointing to a
dynamically generated `constituent' quark mass in
the chiral limit, which we will later estimate.
The function $Z(p)$ was found to
be insensitive to the quark mass.

\subsection{Model fits}
\label{Sec:model_fits}

In order to try to parametrize its behavior, the
mass function $M(p)$ was fitted to the simple
model analytical form
\begin{equation}
aM(pa) = \frac{c}{(ka)^2 + \Lambda^2} + {\muv} \; ,
\label{Eq:model}
\end{equation}
where $k$ was defined in Eq.~(\ref{eq:latticemom}).
We have fitted to data in the window $0\leq ka \leq P$, with $P$
varying between 0.7 and 1.4, in order to verify that the parameters
are insensitive to the fitting window $P$.  Since $S_R$ is far better
behaved than $S_I$ at higher momenta, all the fits have been performed
to the mass function $M(p)$ extracted from our preferred
propagator $S_R$.

\begin{figure}
\begin{center}
\setlength{\unitlength}{1cm}
\setlength{\fboxsep}{0cm}
\begin{picture}(14,7)
\put(0,0){\begin{picture}(7,7)\put(-0.9,-0.4){\ebox{8cm}{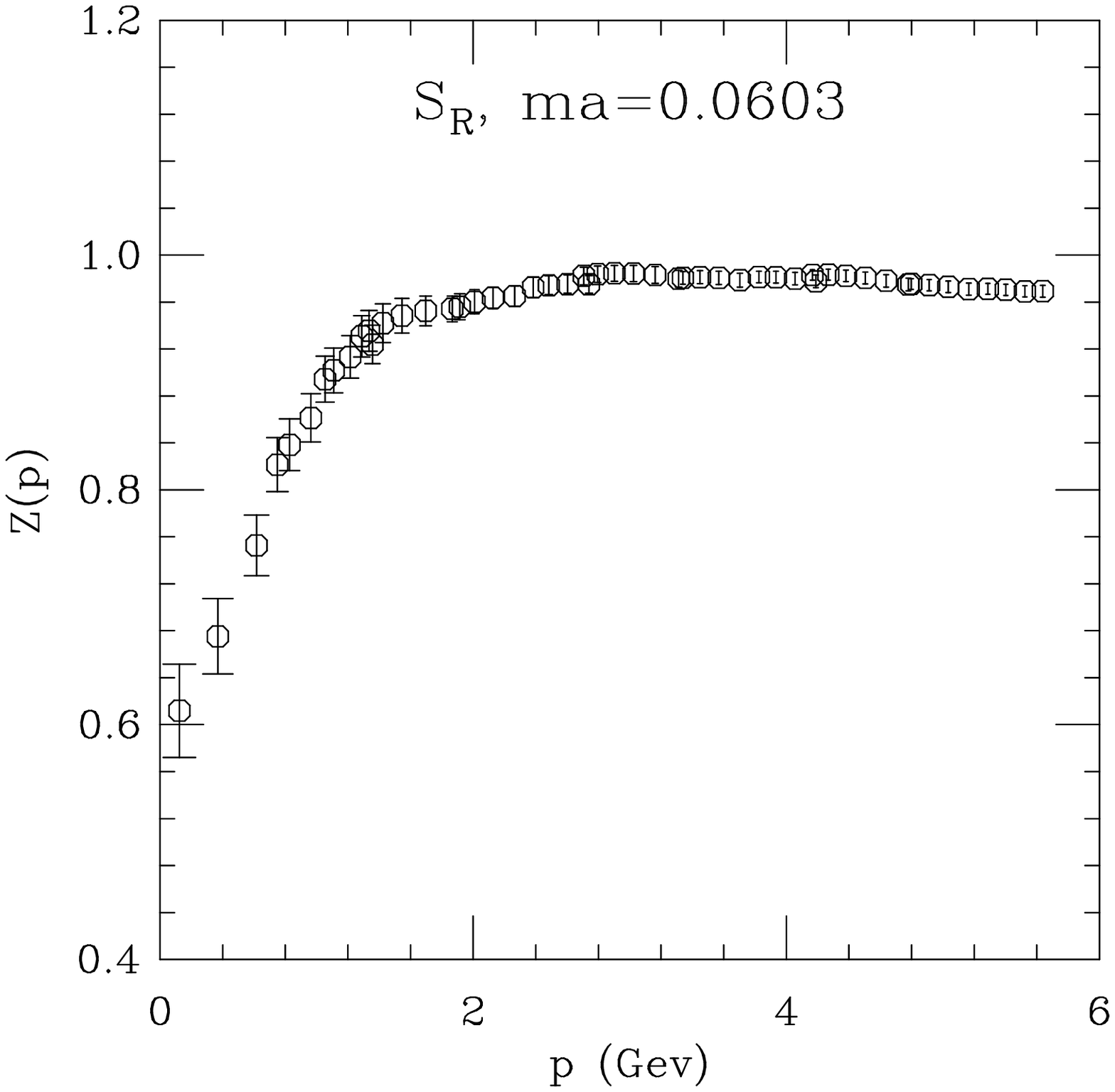}}\end{picture}}
\put(7,0){\begin{picture}(7,7)\put(-0.9,-0.4){\ebox{8cm}{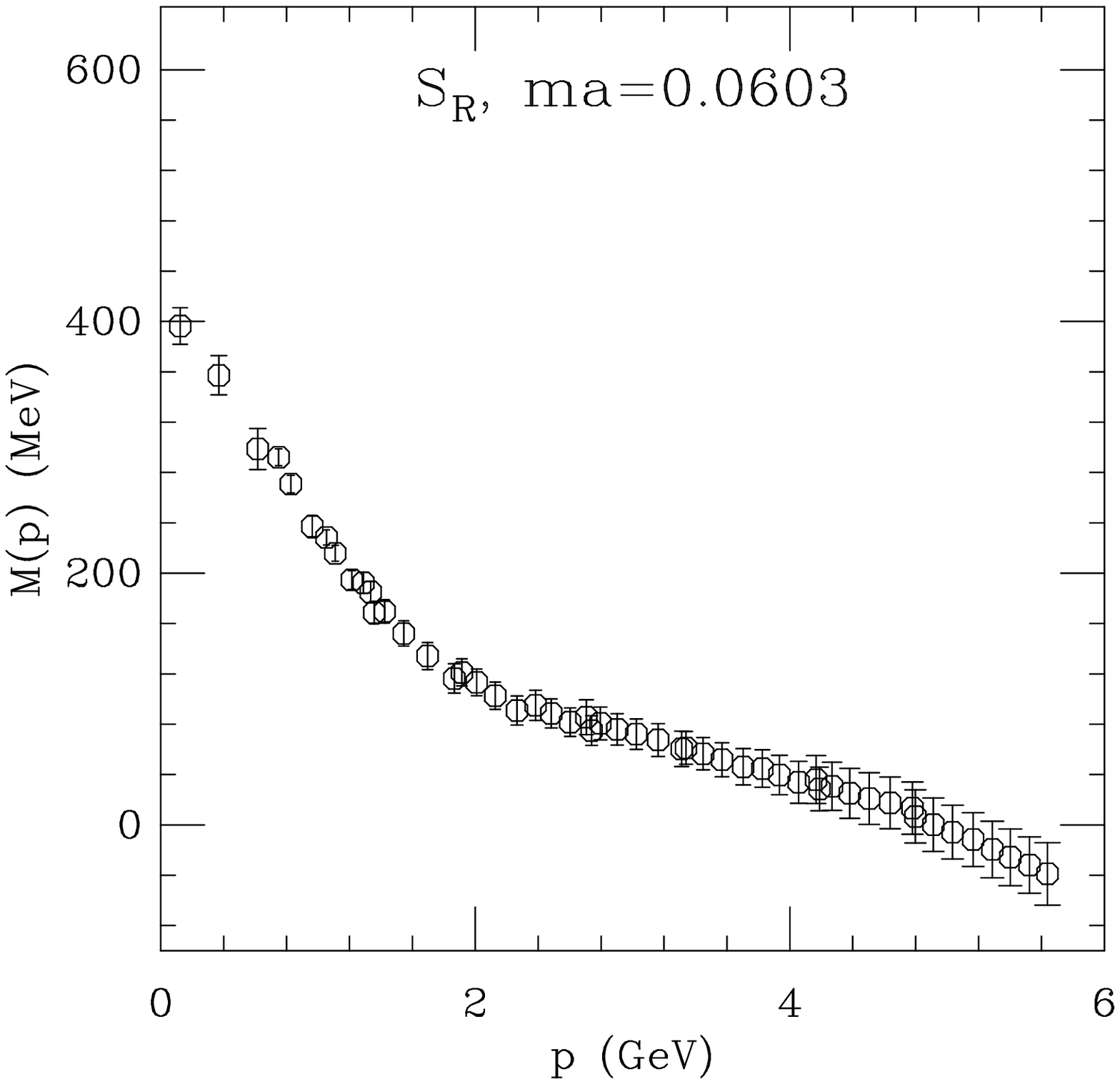}}\end{picture}}
\end{picture}
\end{center}
\caption{ The lattice results for
$Z(p)\equiv Z_2(\mu;a)Z(\mu;p)$ and $M(p)$ for our
preferred form of the quark propagator $S_R(p)$, after both the 
tree-level correction and the cylinder cut.  The vertical scale for
$Z(\mu;p)$ is determined from the above by dividing it by the necessary
renormalization constant (i.e., $Z_2(\mu;a)$) to ensure that
$Z(\mu;\mu^2)=1$.  These are the central results of the studies
reported here.  The bare quark mass used here was $m=112$~MeV and
hence we conclude that $M(p)$ is not
reliable at momenta above approximately $1.5$~GeV.}
\label{Fig:best_Z_and_M}
\end{figure}

The parameter values for $\kappa=0.137$ are shown in table
\ref{tab:fitparams}.  All the fits give a value for $\muv$ which is
consistent with 0.  This is due to the fact that we have not
completely removed these lattice artifacts from $M(p)$ at
intermediate and large momenta.
This indicates that at $\beta=6.0$ with our
preferred improved action and propagator we still do not have
sufficient control of ultraviolet lattice artefacts
that would allow us to extract the ultraviolet running
mass~\cite{Becirevic:2000kb}.  Combining the
fit parameters for all the fits gives a value for the infrared quark
mass of $aM_{\text{ir}} = 0.211\pm 0.008$.

\begin{table}
\caption{Parameter values for best fits to the form of
Eq.~(\protect\ref{Eq:model}) in the window $0\leq ka\leq
ap_{\text{max}}$, for different values of $p_{\text{max}}$, at
$\kappa=0.137$. 
\label{tab:fitparams}}
\begin{tabular}{c|cccc|c}
$ap_{\sss\rm max}$ &$\muv$ &$c$ &$\Lambda$ & $\MIR$
&$\chi^2/N_{\rm{df}}$ \\
\colrule 
0.8 & 0.0017\err{134}{183} & 0.086\err{21}{14} & 0.64\err{6}{4} 
 & 0.210\err{8}{10} & 0.330 \\   
1.0 & -0.0008\err{102}{146} & 0.089\err{19}{13} &
0.65\err{5}{4} & 0.210\err{8}{9} & 0.270 \\
1.2 & 0.0056\err{94}{93} & 0.081\err{10}{8} &
0.63\err{3}{3} & 0.211\err{9}{8} & 0.253 \\
1.4 & 0.0086\err{76}{67} & 0.077\err{8}{7} &
0.61\err{2}{2} & 0.212\err{8}{7} & 0.195
\end{tabular}
\end{table}

\subsection{Infrared quark mass}
\label{Sec:IR_mass}

A quantity of intrinsic interest is the mass function $M(p)$ at zero
momentum in the `chiral' limit $m\to 0$.  This gives a measure of the
dynamical chiral symmetry breaking in the system, and is related to
the order parameter of dynamical chiral symmetry breaking, the chiral
condensate $\bra\psibar\psi\ket$, as well as to the concept of
the `constituent quark mass' used as input in various quark models.

\begin{figure}
\begin{center}
\setlength{\unitlength}{1cm}
\setlength{\fboxsep}{0cm}
\begin{picture}(8,8)
\put(-0.5,-0.4){\ebox{9cm}{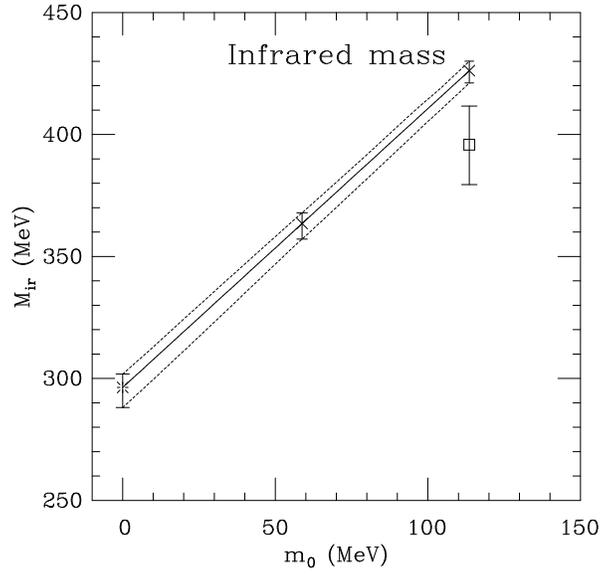}}\end{picture}
\end{center}
\caption{
The infrared value of the quark mass function
$M_{\rm ir}\equiv M(p=0)$, obtained by extrapolating
$M(p)$ to $pa=0$ for two different bare quark masses,
$m= 57.5$ and 112~MeV.
The crosses denote the two values obtained from $S_I$ from 300
configurations and two bare quark masses, while the square is the
value obtained from $S_R$ for a single quark mass.
The burst indicates the chirally extrapolated value of
$M_{\rm ir}$ obtained by a simple straight line interpolation.
}
\label{Fig:M_IR}
\end{figure}

Since we have only computed $S_R$ for one value of the quark mass, we
must use the data from $S_I$ to perform the extrapolation to $m=0$.
Recall that at low momenta the two actions give consistent results
for $M(p^2)$ within errors. 
The results are shown in Fig.~\ref{Fig:M_IR}.  The data point from
$S_R$ at $\kappa=0.137$ is also shown, giving an indication of systematic
uncertainties.  We find a value for $M(0;m=0)$ of $298\pm8\pm30$ MeV,
where the second set of errors is an estimate of the systematic
uncertainty coming from the difference between $S_R$ and $S_I$.
This is reasonably consistent with commonly used values for the
constituent quark mass.

\subsection{Finite volume effects}
\label{Sec:finite_vol}

To determine whether the infrared suppression of $Z(p)$ is real or a
finite volume effect, we can look for anisotropy in the infrared.
Since the temporal extension of the lattice is three times the spatial
extension, the finite volume will affect spatial momenta differently
from timelike momenta, giving an indication of the size of
(anisotropic) finite volume effects.

\begin{figure}
\begin{center}
\setlength{\unitlength}{1.1cm}
\setlength{\fboxsep}{0cm}
\begin{picture}(8,8)
\put(-0.5,-0.5){\ebox{9cm}{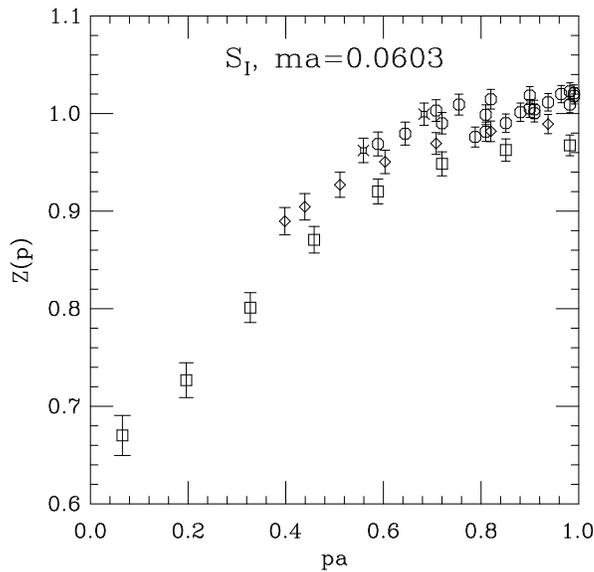}}\end{picture}
\end{center}
\caption{
$Z(p)$ from from the propagator $S_I(p)$, from 70 configurations
at $\kappa=0.137$, i.e., $m=112$~MeV.  The
squares denote purely timelike momenta, while the diamonds denote
points with one unit of spatial momentum.  The fancy squares are
points with half a unit of timelike momentum (ie, nearly purely
spatial momentum).  These data appear to indicate that the infrared
suppression of $Z(p)$ is not a finite volume effect.  
}
\label{Fig:Z_IR}
\end{figure}

Figure~\ref{Fig:Z_IR} shows the infrared behavior of $Z(p)$, with
momenta in different directions plotted separately.  We see that the
finite volume anisotropy, although not negligible, is not sufficient to
explain the infrared suppression.  This indicates that the
suppression is either due to isotropic finite volume effects, or is
a real physical phenomenon.  In model Dyson-Schwinger equation
studies~\cite{Roberts:1994dr}
the dynamically generated quark mass is typically
associated with a dip in $Z(p)$ similar to what is seen here.
There is no discernible anisotropy in the data for the mass function
$M(p)$ at low momenta.  We therefore conclude that finite volume
effects for $M(p)$ are almost certainly negligible.

\section{Conclusion and further work} 
\label{Sec:conclusions}

We have presented initial lattice results for the momentum-dependence
of the quark propagator after implementing a tree-level
correction procedure.
At high momenta, quarks are asymptotically free and so the quark
propagator approaches its tree-level behavior.  We make use of this
fact to subtract off and factor out the tree-level behavior,
replacing it with what should be a more continuum-like
medium and high momentum behavior of the quark propagator.
This approach can only work reliably when the tree-level 
finite-$a$ effects are not too large, i.e., when the tree-level
propagator corresponding to the action of interest is
reasonably behaved at medium and high momenta.
The tree-level correction was seen to dramatically improve the data
for the preferred definition of the improved quark propagator
$S_R$.  The relatively poor behavior of the
tree-level corrected $S_I$ is due to the large tree-level
finite-$a$ effects which require fine tuning to subtract off
correctly.  The unimproved propagator was seen to be unreliable even
at low momenta ($pa\sim 0.4$) and so cannot be trusted even after
tree-level correction.

Although the ultraviolet behavior of the quark propagator is clearly
improved, it remains an open question whether there exists a momentum
window where the lattice data are reliable and perturbation theory is
valid.  One way of checking this would be to calculate the propagator
at three or more different quark masses and attempt a chiral extrapolation
of the mass function in the intermediate momentum region.  If perturbation
theory is valid in this region, the mass function should extrapolate
to zero.

Residual lattice artifacts may also be investigated by studying the
chiral Ward identity \cite{Capitani:1998nr,Cudell:1999kf}, which
should be valid at all momenta.  This involves computing the
pseudoscalar vertex, and while it falls outside the scope of this initial
study, it should be included in future studies of the quark propagator.
In Ref.~\cite{Capitani:1998nr} the Ward identity has been studied for
a slightly different action to ours, and verified for momenta up to
$pa\lesssim1$. 

The central results of this work are summarized in
Figs.~\ref{Fig:best_Z_and_M}, \ref{Fig:M_IR} and \ref{Fig:Z_IR}.
Fig.~\ref{Fig:best_Z_and_M} represents the best estimate from our
currently available data
of the nonperturbative behavior of the quark propagator and is based
on our preferred quark action corresponding to the $S_R$ propagator.
Fig.~\ref{Fig:M_IR} is our extraction of the dynamically generated
infrared or `constituent-like' quark mass.  Finally Fig.~\ref{Fig:Z_IR}
gives an indication of the magnitude of finite volume effects in
$Z(p)$ compared to the nonperturbative effects.

For $pa\lesssim 1$, we see that $M(p)$ falls off with $p$ as
expected.  The values obtained from $S_R$ and from $S_I$ are
consistent, while those for the unimproved propagator $S_0$ differ
significantly.  The infrared mass $M(0)$, which can be thought of
as analogous to a `constituent quark mass', appears to approach a
value of $298\pm8\pm30$ MeV in the chiral limit.

We also find a significant dip in the value for $Z(p)$ at low
momenta.  It must be remembered that the curves for $Z(p^2)$
for different actions need to be rescaled to agree at some
``safe'' low momentum renormalization point before comparing
them.  The finite volume anisotropy is much smaller than the
apparent infrared suppression.  We cannot explicitly
rule out large {\it isotropic} finite volume errors, although based on
experience with earlier gluon propagator studies this seems unlikely.
However, a larger volume is needed to completely resolve this issue.

Since in this inital study, we have used the
mean-field improved value for the clover coefficient $c_{sw}$, and
tree-level improvement for the fermion fields, the quark propagator
still has some residual $\order(a)$ errors as well as $\order(a^2)$
and higher order errors.  To remove the residual $\order(a)$-errors it
would be necessary to compute the non-perturbative values for the
coefficients $b'_q, c'_q$ and $c_n$.  Repeating these calculations at
a different lattice spacing and with other improved quark actions is
also essential to get reliable results for the quark propagator and in
particular for the quark mass function at medium to high momenta.
These studies are currently underway.

\begin{acknowledgments} 

We thank Derek Leinweber for stimulating discussions.  Financial
support from the Australian Research Council is gratefully
acknowledged.  The study was performed using UKQCD data obtained using
UKQCD Collaboration CPU time under PPARC Grant GR/K41663.

\end{acknowledgments}

\appendix

\section{Tree-level expressions --- details}
\label{Sec:tree_level_exp_details}

The dimensionless Wilson fermion propagator at tree level is
\begin{equation}
S_0^{(0)}(p) = \frac{-i\kslash a + ma + \half\khat^2 a^2}
{k^2 a^2 + \left(ma+\half\khat^2 a^2\right)^2} 
 \equiv \frac{1}{D}\left(-i\kslash a + ma + \half\khat^2 a^2\right) \, .
\end{equation}
Since the SW term is proportional to the gauge field tensor, it
vanishes at tree level, so this expression also holds true for the SW
action. 
The tree-level `improved' propagator $S_I$ is given by
\begin{eqnarray} 
S_I^{(0)}(p) & = & (1+ma)S_0^{(0)}(p)-\half  \nonumber \\
 & = &
\frac{-i(1+ma)\kslash a + ma + \half m^2a^2 - \frac{1}{8}\khat^4 a^4 +
\half a^4\Delta k^2} {k^2 a^2 + \left(ma+\half\khat^2 a^2\right)^2}
 \nonumber \\
 & \equiv & \frac{1}{D}\left[-i(1+m a)\kslash a + \Bi\right]
\end{eqnarray}
and the inverse propagator is
\begin{eqnarray}
\left(S_I^{(0)}(p)\right)^{-1} & = &
 \frac{D\left[i\kslash a(1+m a) + \Bi\right]}
{k^2 a^2(1+ma)^2 + {\Bi}^2} \\
 & = & \frac{\left[i\kslash a(1+ma) 
  + ma(1 + \half ma) -\frac{1}{8}a^4\khat^4
 + \half a^4\Delta k^2\right]
\left[k^2 a^2 + \left(ma+\half\khat^2 a^2\right)^2\right]}
{k^2 a^2(1+ma)^2 + \left(ma + \half m^2 a^2
 + \half a^4\Delta k^2 - \frac{1}{8}a^4\khat^4\right)^2} \nonumber
\end{eqnarray}
If we write
\begin{equation}
\left(S_I^{(0)}(p)\right)^{-1} = i\kslash a A^{(0)}(p) + B^{(0)}(p) \, ,
\end{equation}
we find
\begin{eqnarray}
A_I^{(0)}(p) & = &
 -\frac{i}{4N_c}\Tr\left[\kslash a\left(S_I^{(0)}(p)\right)^{-1}\right]/k^2 a^2
 = 1 + \frac{a^2}{4}\frac{\khat^4-m^4}{k^2+m^2} + \order(a^4) 
\label{eq:Aimp-expand} \\
B_I^{(0)}(p) & = & \frac{1}{4N_c}\Tr\left(S_I^{(0)}(p)\right)^{-1} 
 = ma\left(1-\frac{ma}{2} + 
\frac{m^2a^2}{4}\frac{2k^2+m^2+\khat^4/m^2}{k^2+m^2}
\right) + \order(a^4)
\label{eq:Bimp-expand}
\end{eqnarray}

If we write the propagator according to Eq.~(\ref{eq:zmdef}),
\begin{equation}
\left(S_I^{(0)}(p)\right)^{-1} = \frac{1}{\zz_I(p)}
 \left[ i\kslash a + ma + a\dmz_I(p)\right] \, ,
\end{equation}
we find
\begin{eqnarray}
\zz_I(p) & \equiv & \frac{1}{A_I^{(0)}(p)} 
  =  \frac{k^2 a^2(1+ma)^2 + {\Bi}^2}
{(1+ma)D}
\label{Eq:z0-imp}
 \\
a\dmz_I(p) & \equiv & \zz_I(p)B_I^{(0)}(p) - ma 
  =  -\half\frac{m^2 a^2 - a^4\Delta k^2 + a^4\khat^4/4}{1+ma}
\label{Eq:dm0-imp}
\end{eqnarray}
where
\begin{equation}
\Bi \equiv ma+\frac{m^2a^2}{2}+\frac{a^4\Delta k^2}{2}
 -\frac{a^4\khat^4}{8} \, .
\end{equation}

We have defined the rotated propagator $S_R(x,y)$ as
\begin{equation}
S_R(x,y) = (1+\frac{ma}{2})\left(1-\frac{1}{4}\Dslash(x)\right)
 S_0(x,y)\left(1+\frac{1}{4}\Dslashbak(y)\right) .
\end{equation}

At tree level, the fourier transform of this is
\begin{eqnarray}
S_R^{(0)}(p) & = & \left(1+\frac{ma}{2}\right)
\left(1-\frac{ia\kslash}{4}\right)S_0^{(0)}(p)
\left(1-\frac{ia\kslash}{4}\right) \nonumber \\
 & = & \frac{(1+ma/2)}{D}\left(1-\frac{ia\kslash}{4}\right)
   \left(-ia\kslash+m+\khat^2a^2/2\right)\left(1-\frac{ia\kslash}{4}\right)
 \nonumber \\
 & = & \frac{(1+ma/2)}{D}\left[-ia\kslash\Bigg(1+\frac{ma}{2}
    +\frac{3}{16}k^2a^2
    +\frac{1}{4}a^4\Delta k^2\right) + ma \nonumber \\
 &   & -\frac{1}{16}a^3mk^2
    -\frac{1}{32}a^4k^2\khat^2+\half a^4\Delta k^2\Bigg] \nonumber \\
 & \equiv & \frac{1+ma/2}{D}\left(-ia\kslash A'_R(p) + B'_R(p)\right)
\label{eq:tree-rot}
\end{eqnarray}

We can then write
\begin{equation}
S^{(0)}_R(p)^{-1} = \frac{D}{(1+ma/2)D_R}\left(ia\kslash A'_R +
B'_R\right) \, ; \qquad
D_R = k^2 {A'_R}^2 + {B'_R}^2
\end{equation}
>From this we find the expressions for $\zz_R \equiv 1/A_R^{(0)}$ and
$a\dmz \equiv \zz_R B_R^{(0)} - ma$, via
\begin{eqnarray}
A_R^{(0)}(p) & = & \frac{DA'_R}{(1+ma/2)D_R} 
  = 1 + \frac{k^2a^2}{16} + \order(a^2) 
\label{Eq:a0-rot} \\
B_R^{(0)}(p) & = & \frac{DB'_R}{(1+ma/2)D_R} 
 = ma\left(1-\frac{ma}{2} 
 + \frac{m^2a^2}{16}\frac{k^2+4m^2-3k^4/m^2}{k^2+m^2}\right) 
 + \order(a^4)
\label{b0-rot}
\end{eqnarray}
Comparing these expressions with those of Eqs.\ (\ref{eq:Aimp-expand})
and (\ref{eq:Bimp-expand}), we clearly see that the tree-level
$\order(a^2)$ errors in $S_R$ are much smaller than for $S_I$.

\end{document}